\pdfoutput=1
\documentclass[a4paper,twoside]{article}
\usepackage{epsfig}
\usepackage{epstopdf}
\usepackage{subcaption}
\usepackage{calc}
\usepackage{amsmath,amsfonts}
\usepackage{amsmath}
\usepackage{multicol}
\usepackage{pslatex}
\usepackage{apalike}
\usepackage[algo2e]{algorithm2e}
\usepackage[bottom]{footmisc}
\usepackage{placeins}
\usepackage{bm,amssymb}
\usepackage{booktabs} 
\usepackage{xcolor}
\usepackage{tabularx}
\usepackage{algorithm}
\usepackage{algorithmic}
\usepackage{float}
\usepackage{SCITEPRESS}     
\usepackage{mathtools} \usepackage{xspace} \newcommand{\FLIP}{\protect\reflectbox{F}LIP\xspace}

\begin{document}

\title{Deep image-based Adaptive BRDF Measure\vspace{-10pt}}
\author{\authorname{Wen Cao}
Media and Information Technology,Department of Science and Technology, Linköping University, SE-601 74 Norrköping, Sweden\\
\email{wen cao@liu.se}
}


\abstract{Efficient and accurate measurement of the bi-directional reflectance distribution function (BRDF) plays a key role in 
high quality image rendering and physically accurate sensor simulation. However, obtaining the reflectance properties of a material is both time-consuming and challenging. This paper presents a novel method for minimizing the number of samples required for high quality BRDF capture using a gonio-reflectometer setup. Taking an image of the physical material sample as input a lightweight neural network first estimates the parameters of an analytic BRDF model, and the distribution of the sample locations. In a second step we use an image based loss to find the number of samples required to meet the  accuracy required. 
This approach significantly accelerates the measurement process while maintaining a high level of accuracy and fidelity in the BRDF representation.}
\onecolumn \maketitle \normalsize
\setcounter{footnote}{0} \vfill
\section{\uppercase{Introduction}}
\label{sec:introduction}

\begin{figure*}[ht]
 \includegraphics[width=\textwidth]{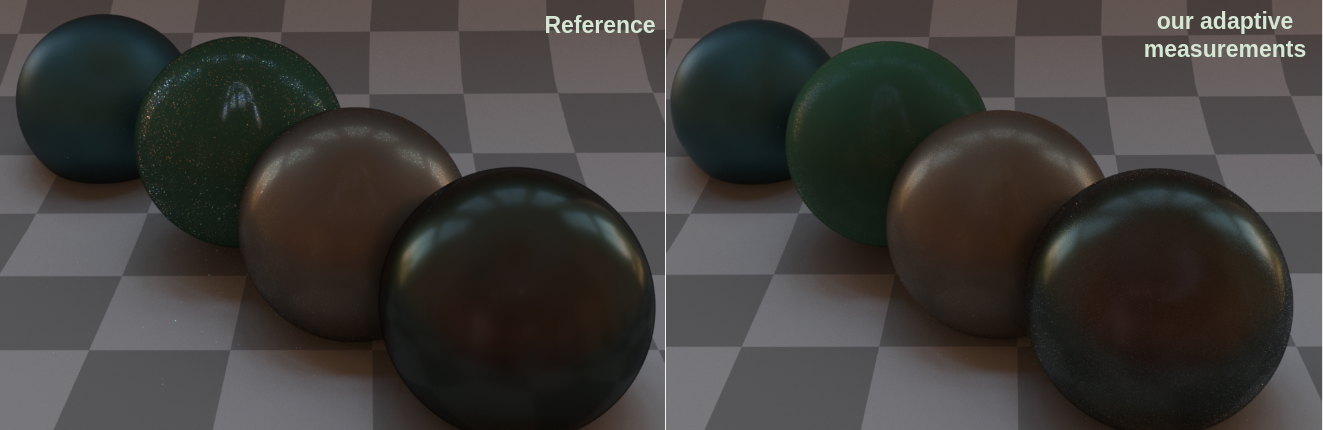}\vspace{-0 mm}
\caption{Rendered balls with ground truth vs our adaptive measurements}
  \label{Fig_Dia}
   \end{figure*}

The bidirectional reflectance distribution function (BRDF), is a fundamental concept in computer graphics, representing the interaction of light with a materials. It is a four-dimensional function that defines the relationship between incoming and outgoing light directions at a material. BRDFs can be represented either by analytic models or by tabulated measurements for every pair of incident and outgoing angles, with each approach having its own advantages and disadvantages. Capture of real, physical BRDFs is an important tool in many applications ranging from photo-realistic image synthesis and predictive appearance visualization in e.g. additive manufacturing  to accurate sensor simulation and modeling of scattering behaviours in industrial processes. However, detailed BRDF measurement is a time-consuming process because it typically requires dense mechanical scanning of light sources and sensors across the entire hemisphere. Several studies ~\cite{nielsen_optimal_2015,dupuy_adaptive_2018} have been conducted to reduce capture time by taking fewer measurements. Recently, neural approaches~\cite{zhang_nerfactor_2021} have been proposed to represent synthetic BRDFs from images, primarily by estimating the material's BRDF parameters.

The objective of this paper is to accelerate BRDF measurements using gonio-reflectometer setups. To incorporate prior knowledge of the material sample, our method uses a small neural network that takes an image of the sample as input to estimate the configuration of a small set of sampling directions to enable efficient BRDF measurement. Specifically, we employ an encoder network to estimate the reflectance parameters of analytic BRDF models from the input image, which are used to adapt the BRDF measurement directions. The method leverages both analytic BRDF models and image-based neural decomposition as priors. These two priors are essential for efficiently utilizing small networks to estimate the adaptive sample distribution.
\section{RELATED WORK}
In this section, we review previous work related to BRDF measurement and neural SVBRDF capture.
\subsection{BRDF Measure}
Generally, people use gonioreflectometers to capture the reflectance of realistic materials by controlling mechanical light sources and camera motions.  For example, acquiring the MERL dataset requires densely sampling 180 azimuth angles, 90 elevation angles, and 90 outgoing directions, totaling approximately 1.46 million samples. To accelerate the acquisition, several methods and devices have been developed. Some research focuses on learning sample patterns from BRDF models to reduce the number of sample locations needed by the measurement apparatus.

Nielsen etc.\cite{nielsen_optimal_2015} and Miandji etc.\cite{miandji_frost-brdf_2024} use the measured BRDF Dataset to train basis to linearly reconstruct the full BRDF samples to accelerate the measurements.  Tong\cite{tongbuasirilai_efficient_2017} use a novel prameterization to acceleratly measure the istropic BRDFs within a single 2D slice. However,Jonathan \cite{dupuy_adaptive_2018}used a laser machine to measure the NDF values of materials to adaptive sample the hemisphere domain. Liu\cite{liu_learning_2023-1} ues meta-learning method to optimize the sampling count of different brdf models. 
\subsection{Neural (SV)BRDF Capture}
 Researchers are exploring deep learning methods to develop lightweight approaches for measuring (SV)BRDF values\cite{noauthor_deep_nodate}. Generally, these methods involve training a network to predict (SV)BRDF parameters such as albedo, diffuse reflection, and roughness.

Valentin\cite{deschaintre_flexible_2023} use encoder-decoder network to estimate the normal ,diffuse albedo,and roughness images from phone-captured images. Xiuming\cite{zhang_nerfactor_2021} uses images as input to predict BRDF values for each pixel based on NeRF output. Zhen\cite{zeng_rgbx_2024} use a diffusion framework to decomposite RGB images to normal, albedo,roughness, metallicity and diffuse maps.

We use a small network to accelerate the measure process and validate our method in MERL dataset.
\section{APPROACH}
Typically, BRDF model encompass both analytic functions and empirically measured values. 
Our method employs a Convolutional Neural Network (CNN) encoder to estimate the BRDF parameters of an analytic model from a single image. Subsequently, importance sampling techniques are used to derive an adaptive sampling pattern for the input material.  
\begin{figure*}[h]
  \includegraphics[height=1.7in,width=\textwidth]{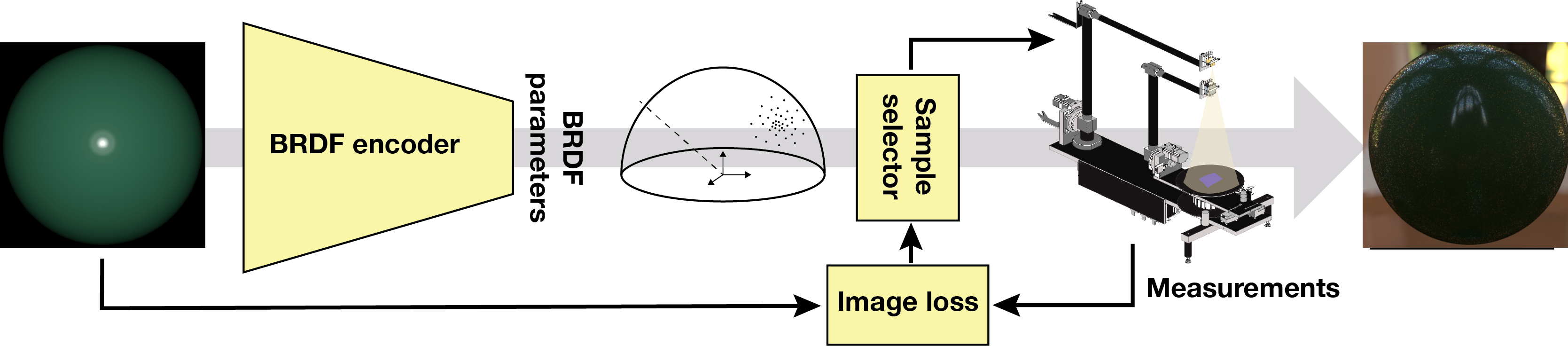}
  \caption{Method Flowchart of Deep Image-based Adaptive Reflectance Measure. For a fixed material, we use its image as input to an encoder network, which then estimates the BRDF parameters of it. An adaptive sampler use these parameters to determine the outgoing direction locations. Finally, we progressively increase the number of these locations to achieve the minimum number of samples required while maintaining high fidelity. }
  \label{fig:teaser}
\end{figure*}

Furthermore, these sampling values are used to render the estimated image, and the image loss is calculated to determine the optimal sample number, as illustrated in Fig.~\ref{fig:teaser}. In this work, we present a novel lightweight approach for sampling the measured BRDF based on an analytic model. To validate the accuracy of our approach, we conduct a virtual acquisition experiment using the MERL database.

\subsection{BRDF Estimation}
\label{net_brdf}
\begin{figure}[h]
  \centering
  \includegraphics[height=1.5in,width=\linewidth]{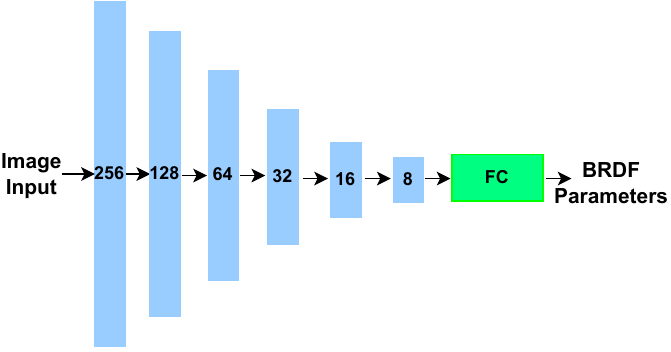}
  \caption{Encode Network Architecture: The blue boxes denote convolutional layers that are integrated with batch normalization and ReLU activation functions. The dimensions of these layers are the numbers inside them. A green box represents a fully connected layer outputting the BRDF parameters.}
   \label{fig:brdfnet}
\end{figure}

Drawing inspiration from deep learning techniques used in SVBRDF capture from images, we train a convolutional neural network to encode isotropic material images into their corresponding appearance parameters for a set of analytical models. We focus on small network architectures and train them using a combination of synthetically generated image and BRDF data, and fine tune with real captured data. The architecture of the BRDF estimation network is illustrated in Fig.~\ref{fig:brdfnet} and employs sigmoid activation functions for the parameter outputs.

\subsubsection{Ward BRDF model}
We use the Ward BRDF model parameterized by its specular $\rho_s=1-\rho_d$ ,roughness $\alpha$ and albedo $\rho_d$ to train the network.
The isotropic Ward BRDF $f_r(\mathbf{i}, \mathbf{o})$ with incoming direction i and outgoing directin o is:
\begin{equation}
f_r(\mathbf{i}, \mathbf{o})=\frac{\rho_d}{\pi}+\frac{\rho_s}{4 \pi \alpha^2 \sqrt{\cos \theta_i \cos \theta_o}} e^{-\frac{\tan ^2 \theta_h}{\alpha^2}}
\label{equ:brdf}
\end{equation}
where  $\theta_i$,$\theta_o$,$\theta_h$ is the incoming angle, outgoing angle, half angle.

To train the encoder network, we use the Loss function of Ward BRDF model as below
\begin{equation}
\mathcal{L}_{\text {loss }}= \|\hat{I}-I\|_1+\|\left(\hat{\rho}_d,\hat{\alpha}\right)-\left({\rho_d,\alpha}\right)\|_1
\label{equ:loss}
\end{equation}
where $I$ represents the rendered image based on the BRDF parameters. L1 loss is used for both images and parameters. For further detailed results, please refer to section ~\ref{all_results}.
\subsubsection{Merl Dataset}
We also use Measured BRDFs Dataset (MERL) to validate this work. We predict alpha $\alpha$ for each image by training from scratch the nerual brdf network.
\begin{equation}
f_r(\mathbf{i}, \mathbf{o})=f_{ggx}^{\prime}(\alpha)
\end{equation}
where  $ f_{ggx}^{\prime}$ is microfacet brdf model to represent the reflectance. We optimize the network to minimize the loss between estimation and ground truth of alpha value and images.
\begin{equation}
\mathcal{L}_{\text {loss }}=\|\hat{I}-I\|_1+\|\left(\hat{\alpha}\right)-\left({\alpha}\right)\|_2
\label{equ:loss2}
\end{equation}
The ground truth alpha values for each MERL material are derived by fitting their respective total BRDF data to microfacet BRDF model~\cite{zhang_nerfactor_2021,zheng_compact_2022}. Based on the BRDF parameters estimated by the network, the estimated images $\hat{I}$ are rendered using the Microfacet BRDF model. We employed an L1 loss between the predicted images $\hat{I}$ and the ground truth images $I$ ,while an L2 loss was applied to the parameter estimates.
\subsection{Adaptive Reflectance Sample}
\label{sampling}
Using the BRDF estimation network, we are able to derive the BRDF parameters from an input image. We draw inspiration from prior work and utilize the inverte BRDF importance sampling  to drive a adaptive sampling distribution of the outgoing hemisphere similar with Jakob's and YAOYI 's research~\cite{bai_bsdf_2023}. This adaptive sampling strategy effectively minimizes measurement time by targeting only those directions specified by the input BRDF material pattern. For instance, in the case of a mirror-like material, sampling is concentrated on the delta regions directly opposite the incoming light direction. In contrast, for diffuse materials, a uniform sampling pattern is implemented throughout the hemisphere. In practice, the BRDF sampling pattern for most materials typically falls between these two extremes.For more detailed explanation of importance sample, we can refer to Bai et al.~\cite{bai_bsdf_2023}.
\begin{figure}[h]
  \includegraphics[height=1.2in,width=0.45\textwidth]{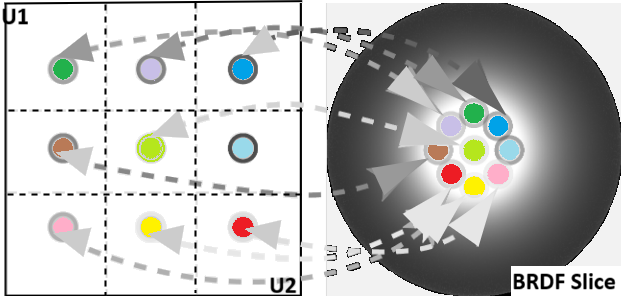}
  \caption{A visualization of the process g in Eq.~\ref{brdf_eqa} to calculate the adaptive sampler's position. We start from get sample points $\left(u_1, u_2\right)$ on a uniform grid in the unit square $[0,1]^2$. The importance sampling process takes a sample point $\left(u_1, u_2\right)$, and maps it to position  on a 2D BRDF slice ,and reverse vise works by its inverse function.}
 
  \label{fig:imp}
\end{figure}

The rendering equation of general BRDF model is as follows:
\begin{equation}
I(\mathbf{i})=\int_{\mathcal{U}^2} f_r\left(\mathbf{i}, g(\mathbf{u})\right) L_i(g(\mathbf{u}))\left\|J_g(\mathbf{u})\right\| \mathrm{d} \mathbf{u}
\label{brdf_eqa}
\end{equation}
$L_i$ is the incident radiance,$I_i$ is the intensity of integration, where $J_g$ is the Jacobian of $g$ .
\begin{equation}
\begin{aligned}
p\left(\boldsymbol{\omega}_{\mathrm{o}}\right)=w_{\mathrm{d}} \cdot p_{\mathrm{d}}\left(\boldsymbol{\omega}_{\mathrm{o}}\right)+w_{\mathrm{S}} \cdot p_{\mathrm{s}}\left(\boldsymbol{\omega}_{\mathrm{o}}\right)
\end{aligned}
\end{equation}
where $w_{\mathrm{d}} +w_{\mathrm{s}} =1$, $p(\omega_o)$ is the PDF of outgoing direction $\omega_o$. The diffuse PDF $p_d$ is a simple cosine-weighted distribution and PDF $p_s$ is the specular distribution based on BRDF specular lobe. The equation $g$ construct the adaptive sampling in the outgoing hemisphere domain according to the specular distribution $p_d$ without the diffuse distribution as Fig.~\ref{fig:imp} and its inverse $g^{-1}$ map the location of BRDF slice back to the unit square which is used to bilinear interpolation to fully evaluate the brdf values in rendering function to produce the image.

Thus,we focus on samplers suitable for representation $f_r$ : an invertible function g from random variates 
$\mathbf{u} \in[0,1]^2$ into outgoing directions $\omega_o$ and its associated probability density function(PDF) 
$p\left(\omega_{\mathrm{o}}, \omega_{\mathrm{i}}\right)$. The shape of $p$ shuld closely matches $f_r$ to achieve low variance.

\subsubsection{Ward Importance Sampling}
For Ward brdf model, the importance sampling equation is as below:
\begin{equation}
\begin{aligned}
\theta_h & =\arctan (\alpha \sqrt{-\log u_1}) \\ 
\phi_h & =2 \pi u_2
\end{aligned}
\label{equ:im_sa}
\end{equation}
where $\phi_h,\theta_h$ is the half angle; $u_1,u_2$ is the uniform variates on $[0,1]^2$. Then,we compute the inverse of the function~\ref{equ:im_sa}, to evaluate the measured brdf values in rendering equation by equation~\ref{equ:re_sa}.
\begin{equation}
\begin{aligned}
u_1 & =e^{-\frac{\tan ^2 \theta_h}{\alpha^2}} \\
u_2 & =\frac{\phi_h}{2 \pi} 
\end{aligned}
\label{equ:re_sa}
\end{equation}
Then, we use $\omega_o=2(\omega_h\cdot \omega_i)\omega_h-\omega_i$ to determines the outgoing direction $\omega_o(\phi_h, \theta_h)$.

\subsubsection{MERL Dataset}
Not similarly to the Ward model, an analytical function for obtaining the probability density function (PDF) values is not available for the MERL dataset. Therefore, we approximate the PDF using the microfacet BRDF model~\cite{dupuy_photorealistic_nodate} and employ its importance sampling method to achieve adaptive measurements. Additionally, we utilize the inverse of this function in the rendering process. Detailed functions and methodologies can be found in~\cite{dupuy_photorealistic_nodate}.


Finally, we adaptive sampled the outgoing direction according to PDF values is employed to get the measurements of BRDF.  These measurements can guide the goniometers, facilitating precise and efficient measurement of the input material, meantime reducing capture time.The incoming directions are uniformly sampled within the cosine-weighted hemisphere.
\section{Implemantation}
We use Mitsuba 3.0~\cite{noauthor_mitsuba_nodate}and its Python bindings to render these $256\times 256$ images and PyTorch to implement the network.

\subsection{Dataset}
To train the neural network of Ward BRDF model,we create a dataset covering the full range of Ward BRDF parameter-$\alpha$. Images with varying roughness and diffuse values were rendered using the Ward BRDF function, a single point light source, and a sphere in Mitsuba. The dataset consists of 4,000 training images and 100 test images.

Similarly, for the MERL dataset, we fit each material to the Microfacet BRDF model. We first create datasets by rendering images with varying alpha and albedo values using the Microfacet BRDF model. These images are used to pretrain the BRDF estimation network, addressing the limited amount of measured data in the MERL dataset. We then fine-tune the BRDF estimation network using the MERL training images, with the fitted alpha values serving as ground truth. The Microfacet dataset comprises 40,000 training images and 100 test images, while the MERL dataset includes 85 training images and 15 test images rendered with different materials. Images from these datasets are provided in the supplementary materials.

\subsection{Estimation}
The BRDF estimation network are depicted in Fig.~\ref{fig:brdfnet} . We train the network in Nivida Geforce RTX 4080 with 15GB memory.

\begin{figure}[h]
  \centering
  \includegraphics[height=1.8in,width=0.48\linewidth]{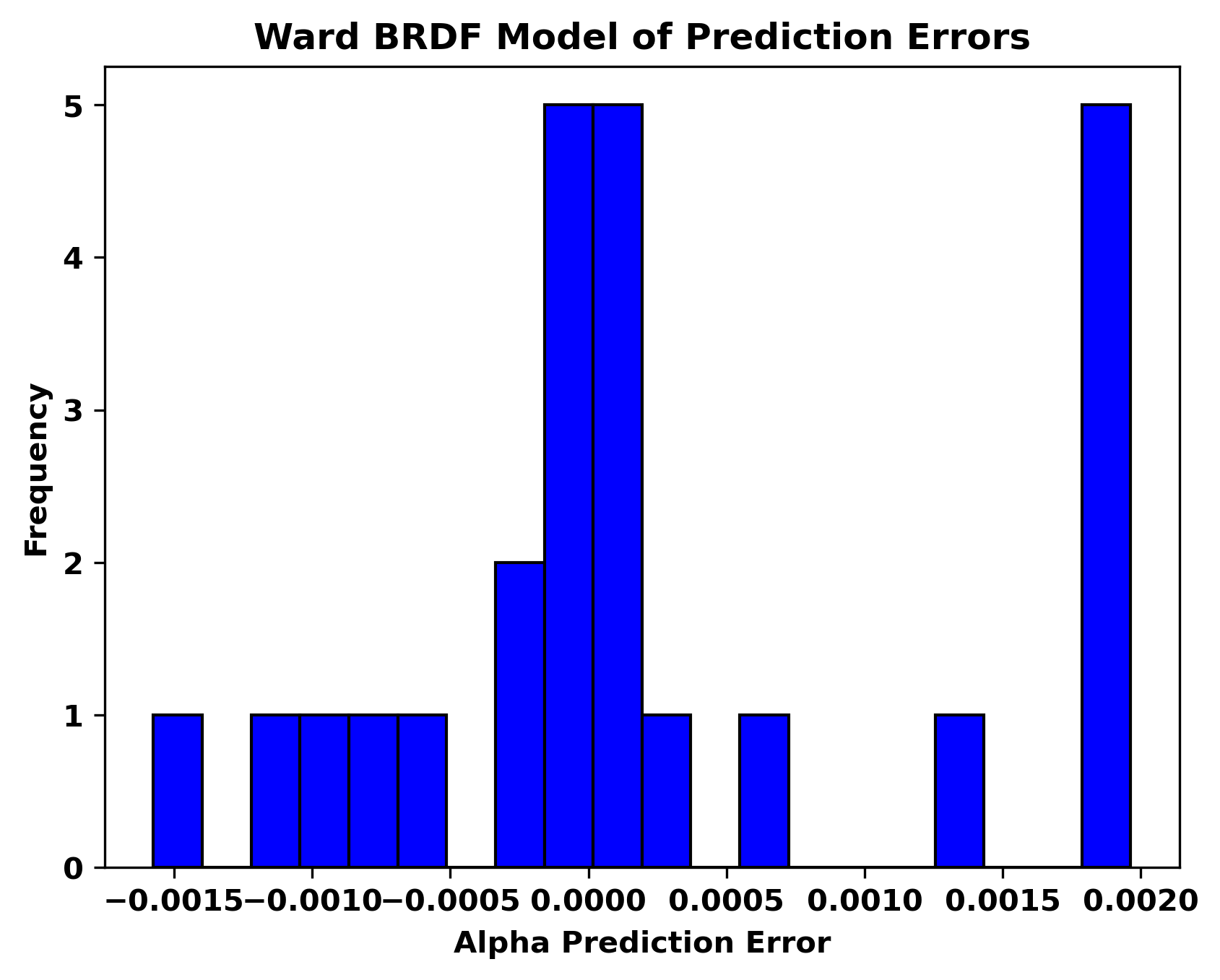}
  \includegraphics[height=1.8in,width=0.48\linewidth]{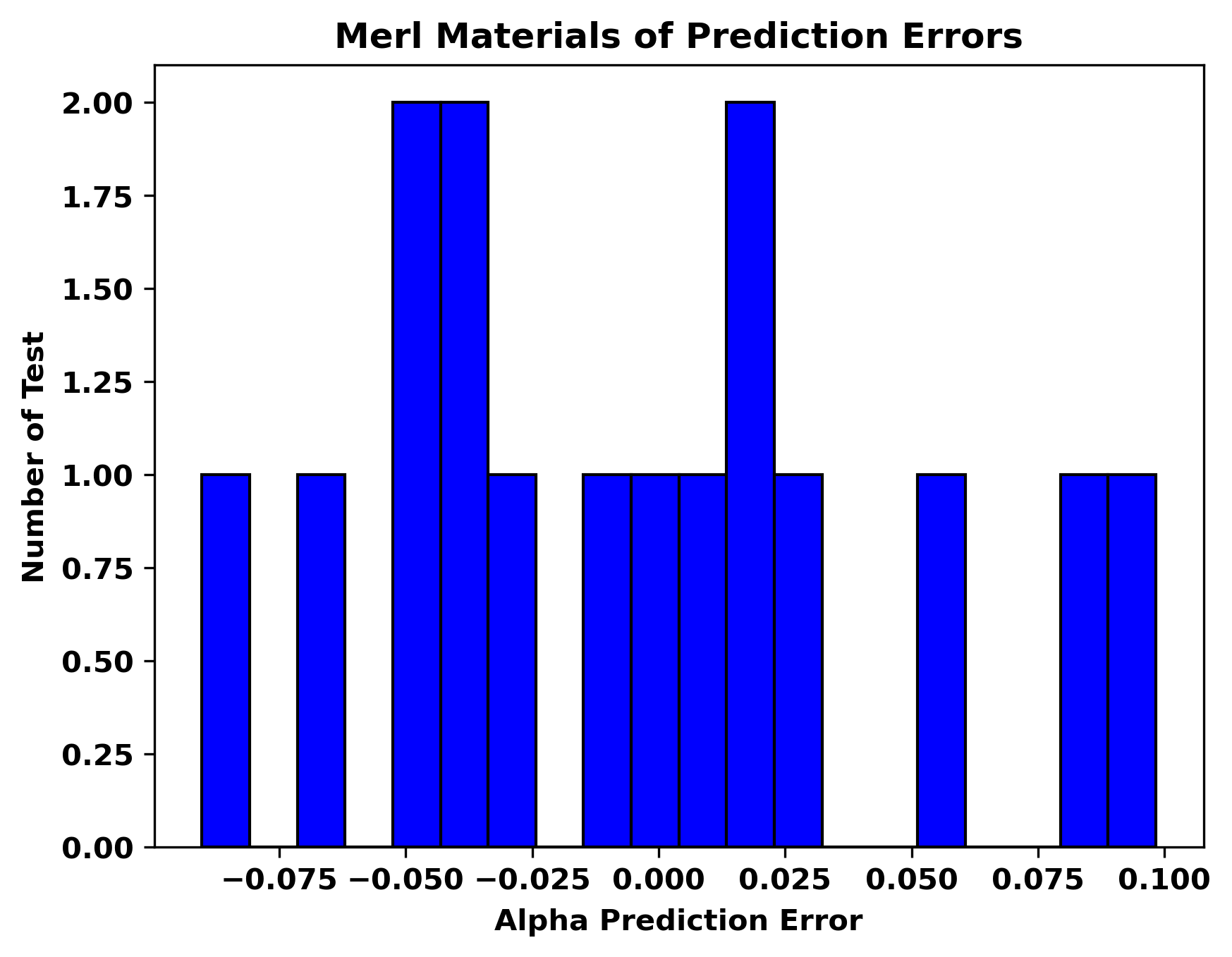}
  \caption{Predicted Values from BRDF Nerual Network VS Ground Truth.}
\label{fig:netresults}
  
\end{figure}

We demonstrate that our BRDF estimation network accurately predicts BRDF parameters for both the Ward model and the MERL dataset, as illustrated in Figure~\ref{fig:netresults}. For the Ward BRDF model, the network achieves nearly perfect predictions. For the 15 materials in the MERL test set, the estimation errors range between –0.1 and 0.1. In most BRDF models, albedo represents the base color, while alpha describes the shape of the BRDF reflection lobe. Since the base color is visually evident in our results, we primarily present the network's alpha estimations in Figure~\ref{fig:netresults}.

\subsection{Sample Count}
\label{sample_num}
\begin{figure*}[ht]
    \setlength{\tabcolsep}{0.02cm}
    \begin{tabularx}{1.30\linewidth}{ccccccccc}
       \includegraphics[height=0.5in,width=0.11\linewidth]{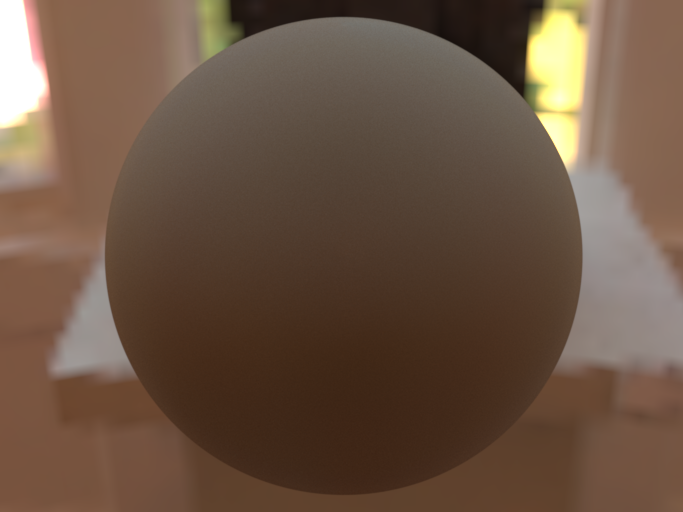} 
       &\includegraphics[height=0.5in,width=0.11\linewidth]{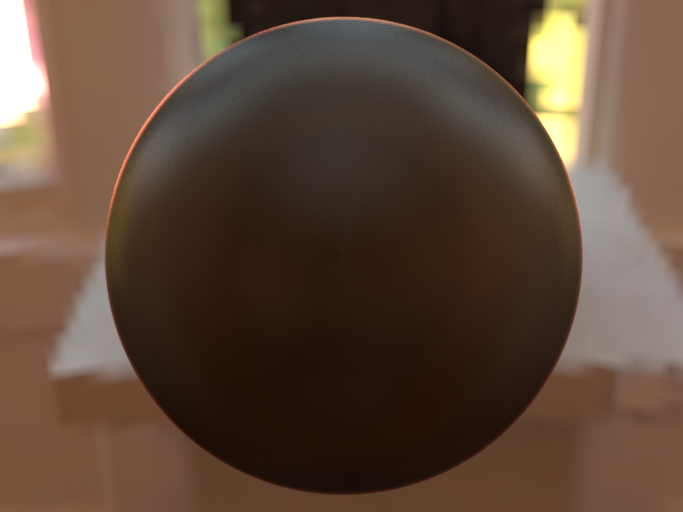} 
        &\includegraphics[height=0.5in,width=0.11\linewidth]{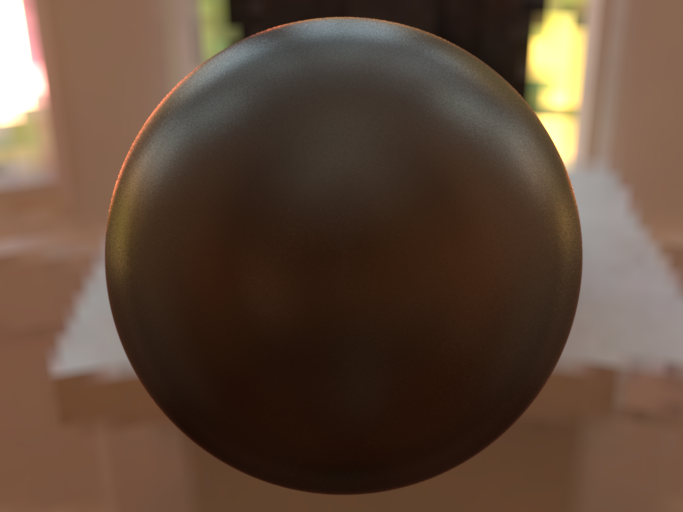} 
        &\includegraphics[height=0.5in,width=0.11\linewidth]{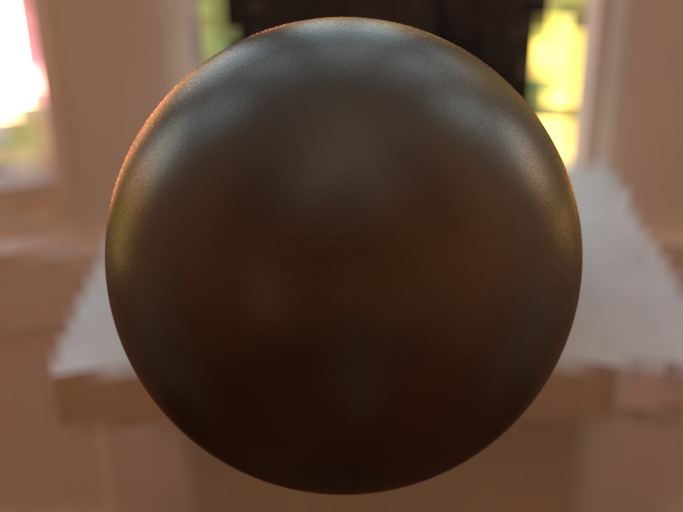} 
        &\includegraphics[height=0.5in,width=0.11\linewidth]{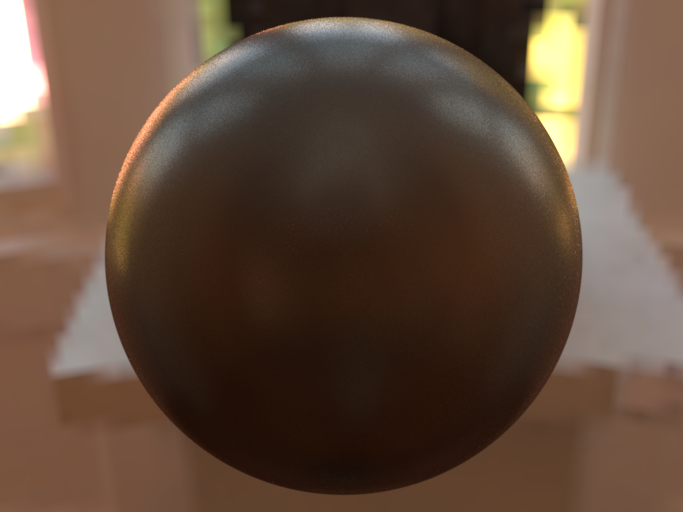} 
        &\includegraphics[height=0.5in,width=0.11\linewidth]{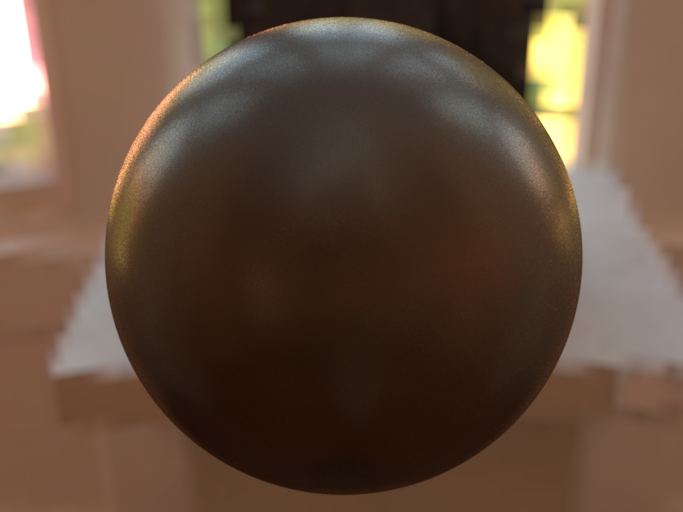} 
        &\includegraphics[height=0.5in,width=0.11\linewidth]{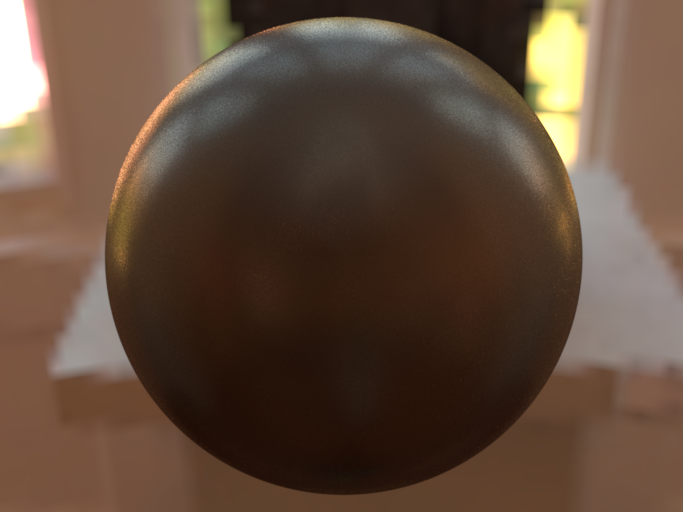} 
        &\includegraphics[height=0.5in,width=0.11\linewidth]{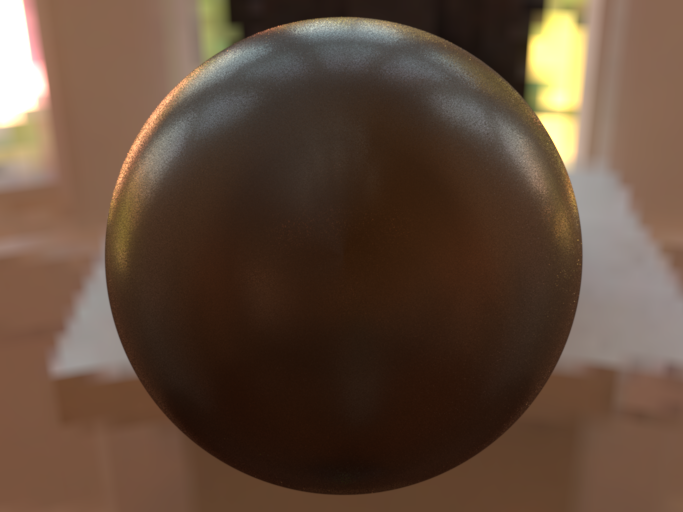}
        &\includegraphics[height=0.5in,width=0.11\linewidth]{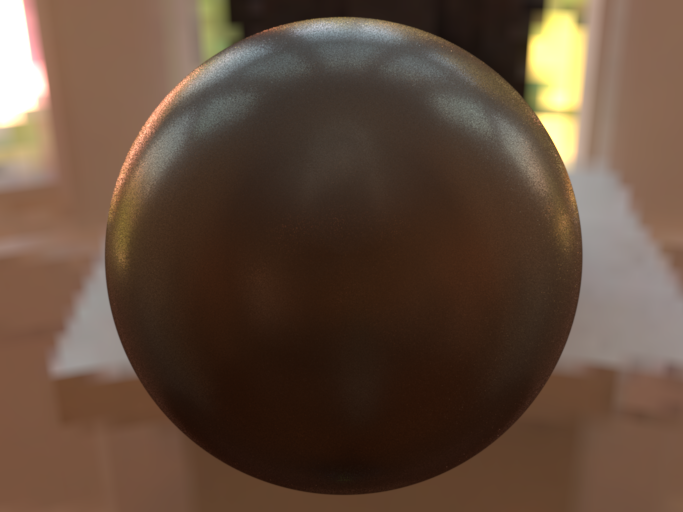}
    \\$2\times2$&$4\times4$&$6\times6$&$8\times8$ &$10\times10$&$12\times12$&$14\times14$&$16\times16$&$18\times18$\\
        \includegraphics[height=0.5in,width=0.11\linewidth]{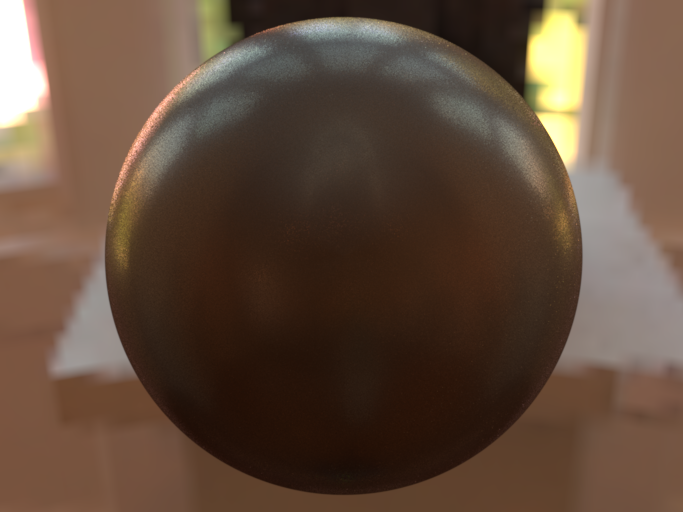}
        &\includegraphics[height=0.5in,width=0.11\linewidth]{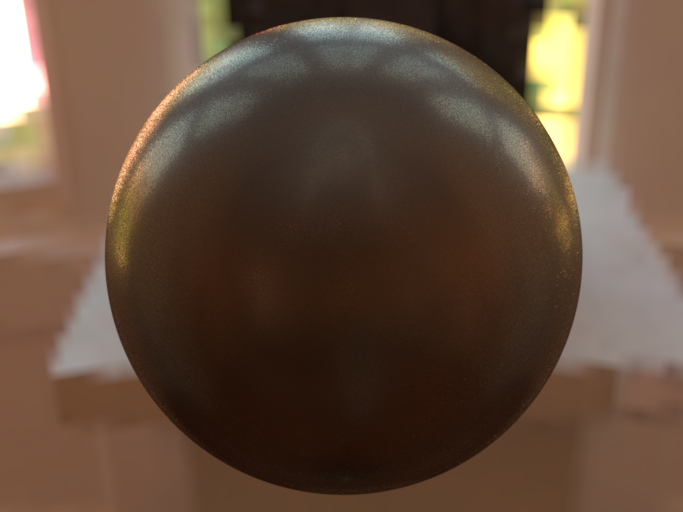} 
        &\includegraphics[height=0.5in,width=0.11\linewidth]{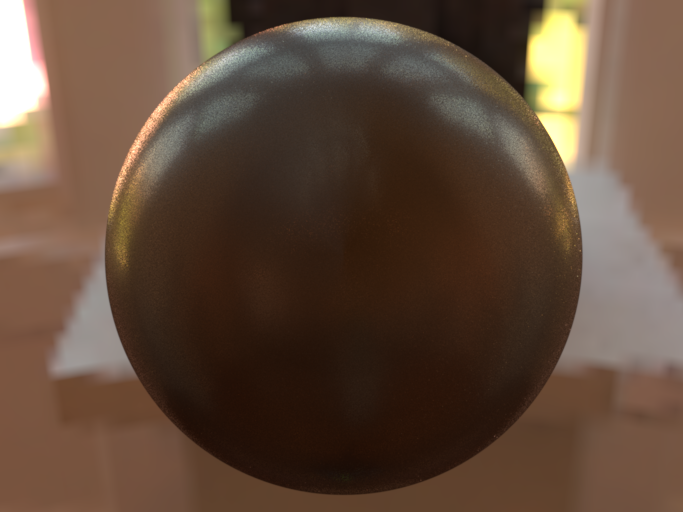} 
        &\includegraphics[height=0.5in,width=0.11\linewidth]{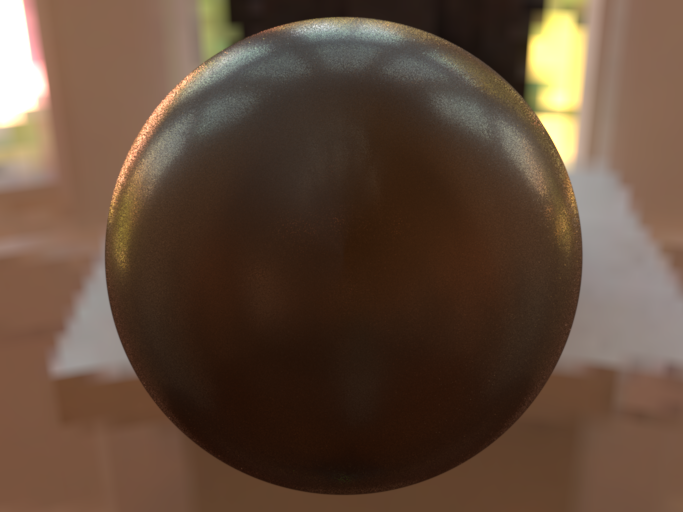} 
        &\includegraphics[height=0.5in,width=0.11\linewidth]{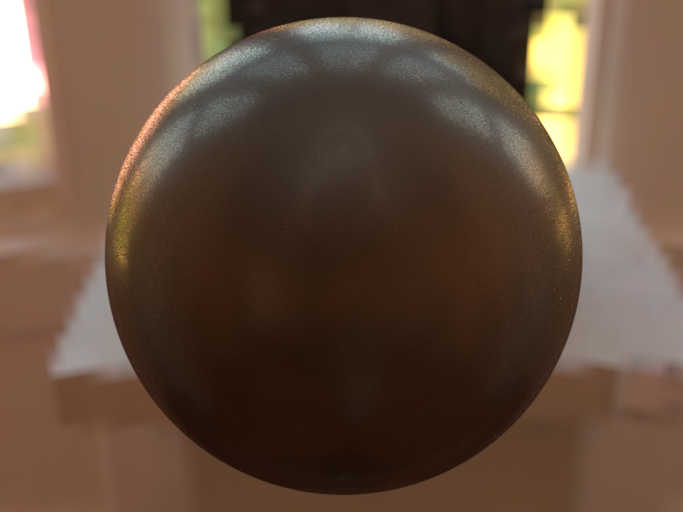} 
        &\includegraphics[height=0.5in,width=0.11\linewidth]{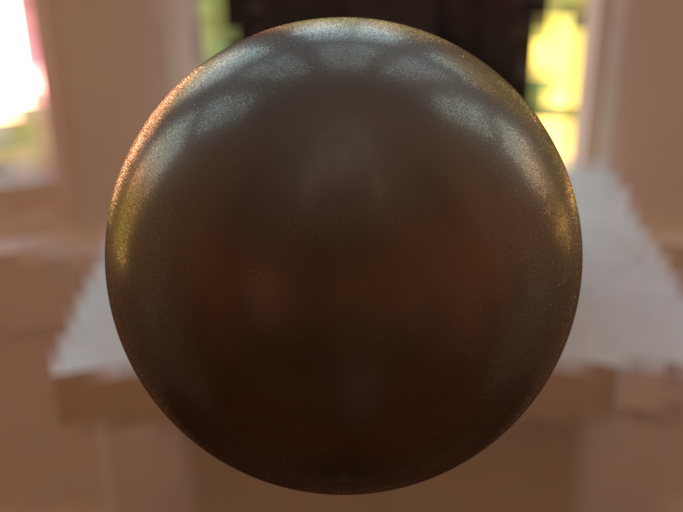} 
        &\includegraphics[height=0.5in,width=0.11\linewidth]{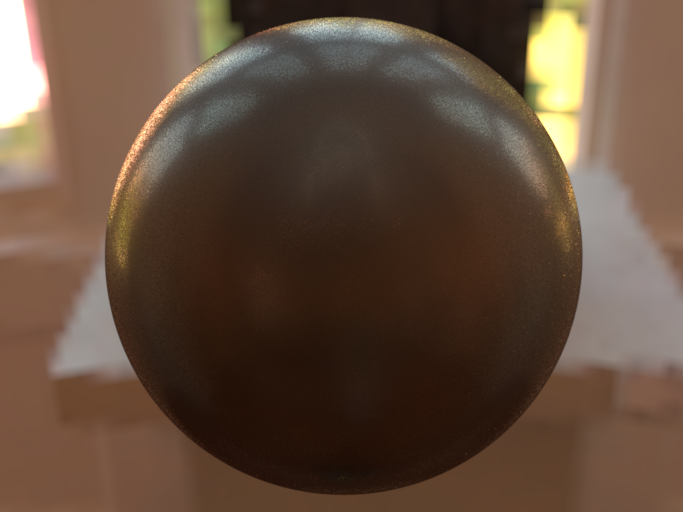} 
        &\includegraphics[height=0.5in,width=0.11\linewidth]{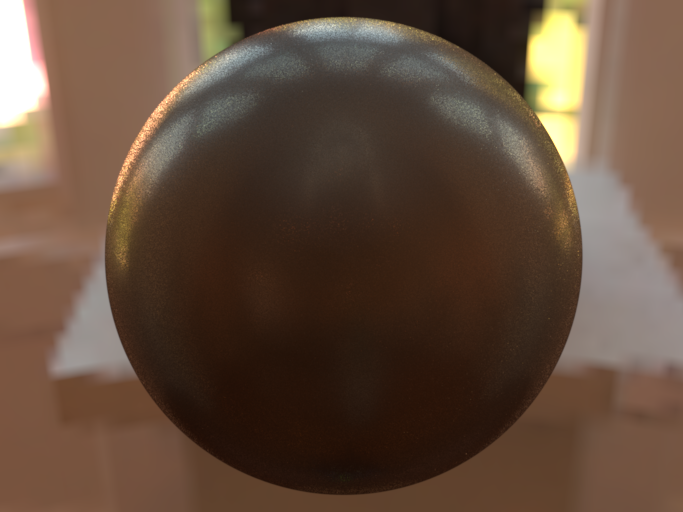}&\includegraphics[height=0.5in,width=0.11\linewidth]{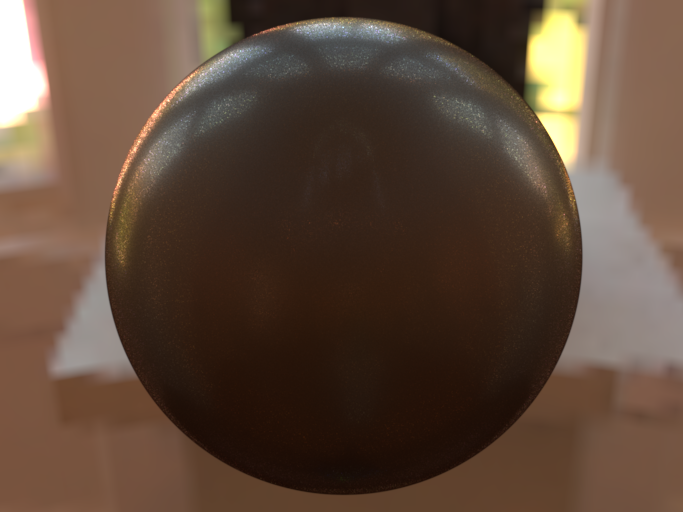} \\
 $20\times20$&$22\times22$&$24\times24$&$26\times26$&$28\times28$&$30\times30$&$32\times32$&$34\times34$&Ground Truth\\
    \end{tabularx}
    \caption{Rendered images from our adaptive measurements using an increased number of samples for material: Aluminum bronze. }
    \label{sample_count}
    \end{figure*}

We treat the number of outgoing directions as hyper parameters of the entire pipeline. For each material, we optimize samples numbers by the image loss between the rendered images by the measurements and the ground truth. In Fig.~\ref{sample_count}, we show rendered images of Aluminum bronze using sample counts ranging from $2\times2$ to $34\times34$. The images demonstrate that performance improves as the number of samples increases. However, once the sample count reaches $16\times16$, performance no longer continues to improve.Therefore, we select $16\times16$ as the final sample count for our measurements of Aluminum bronze .

\begin{figure}[h]
  \centering
  \includegraphics[height=1.8in,width=0.48\linewidth]{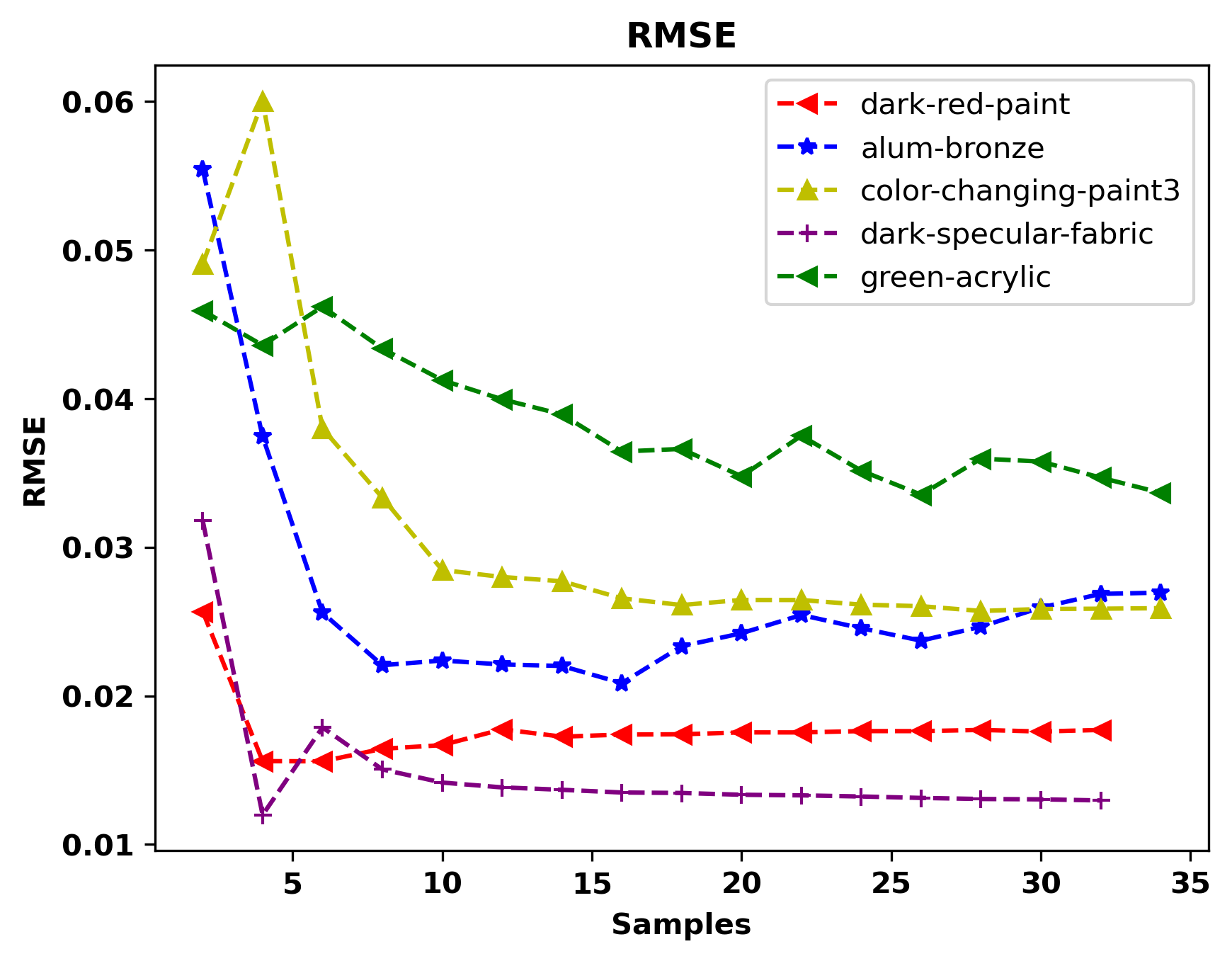}
  \includegraphics[height=1.8in,width=0.48\linewidth]{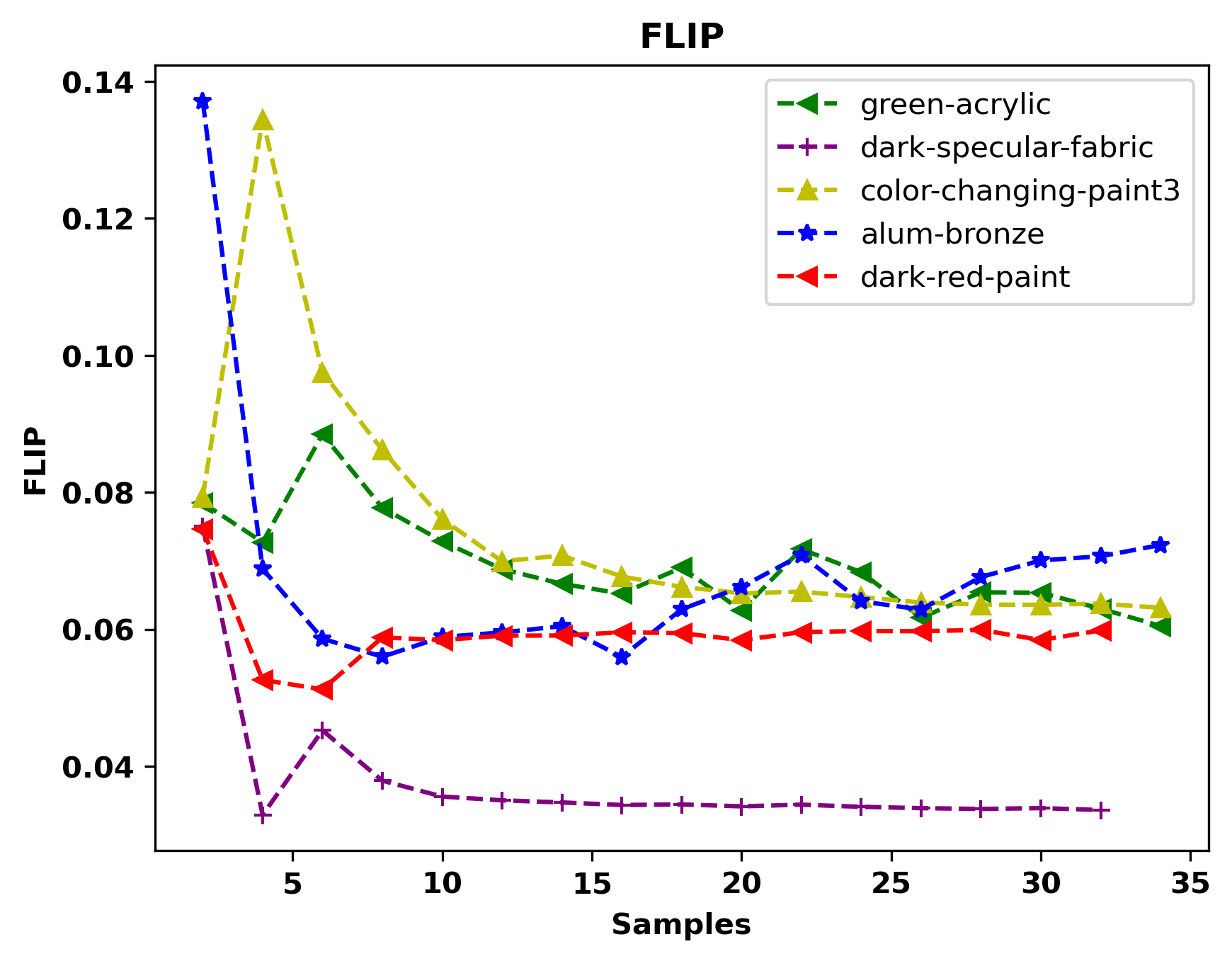}
  \caption{RMSE and Mean $\FLIP$ Error Performance of five different materials with increased samples.}
\label{fig:samplecounts} 
\end{figure}

However we observe that the plot lines vary across different materials, as shown in Fig.~\ref{fig:samplecounts}. Generally, materials with higher specular reflections require more samples. To minimize the size of the measurements, we set the maximum direction count to $32\times 32$, as increasing the number of samples beyond this point yields visually negligible improvements for most materials.

For the number of incident directions, we use one $\phi$ point and eight $\theta$ points sampled from a uniform cosine distribution, adhering to the rotational symmetry of isotropic materials. The number of directions was determined based on previous work in ~\cite{dupuy_adaptive_2018}.

\section{Results}
In this section, we present the qualitative results and comparisons with state-of-the-art methods for BRDF acquisition. For quantitative evaluations, we employ metrics as Peak Signal-to-Noise Ratio (PSNR), Root Mean Square Error (RMSE), and mean $\FLIP$ error~\cite{andersson_flip_2020}. Additionally, we use $\FLIP$ error maps to visualize the errors in the rendered images.

\subsection{Rendered Results}
\begin{figure}[ht]
    \setlength{\tabcolsep}{0.02cm}
    \begin{tabularx}{1.30\linewidth}{cc}
        \includegraphics[height=1.0in,width=0.40\linewidth]{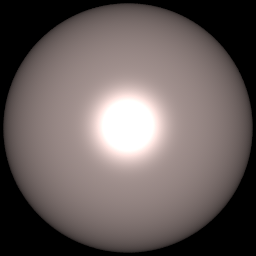} &
        \includegraphics[height=1.0in,width=0.40\linewidth]{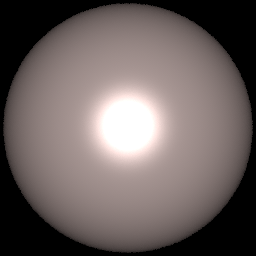} \\
        Ground Truth&Nerual BRDF Prediction\\
        \includegraphics[height=1.0in,width=0.40\linewidth]{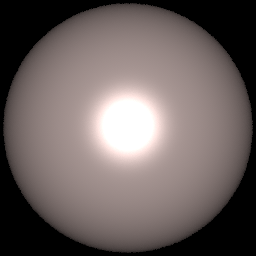} &\includegraphics[height=1.0in,width=0.40\linewidth]{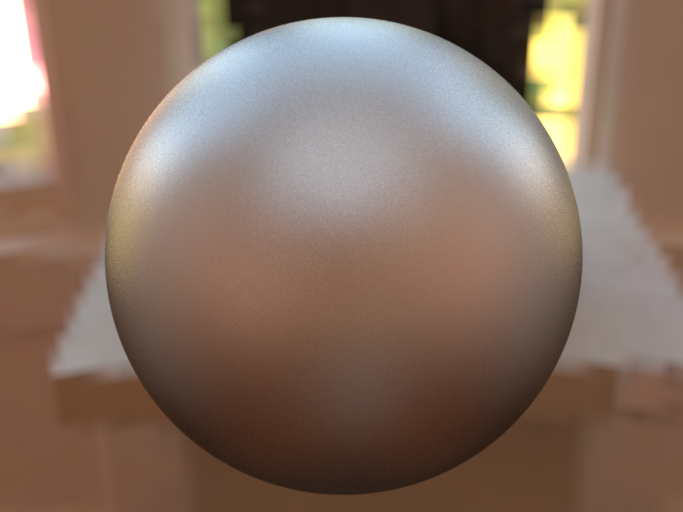} \\
        Render with measurement& Novel Light with measurement
      \end{tabularx}
\caption {Rendered images based on the results of each component of our method with the Ward BRDF model. Note that the Neural BRDF estimation image is rendered using the BRDF parameters estimation by our neural network.}
\label{measure}
\end{figure}

\begin{figure}[ht]
    \setlength{\tabcolsep}{0.02cm}
    \begin{tabularx}{1.30\linewidth}{cc}
    \includegraphics[height=1.0in,width=0.40\linewidth]{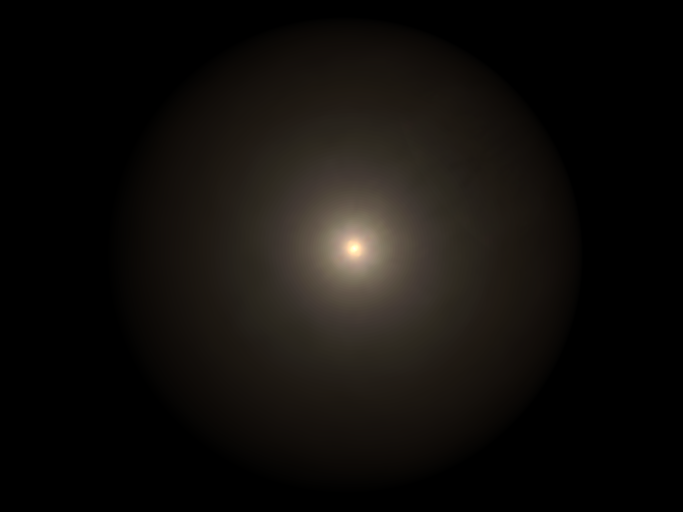} &\includegraphics[height=1.0in,width=0.40\linewidth]{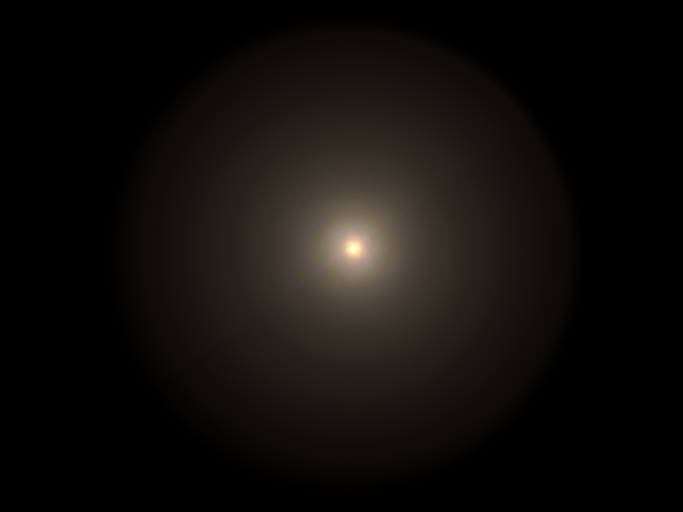}\\
    Ground Truth&Render with measurement\\
    \includegraphics[height=1.0in,width=0.40\linewidth]{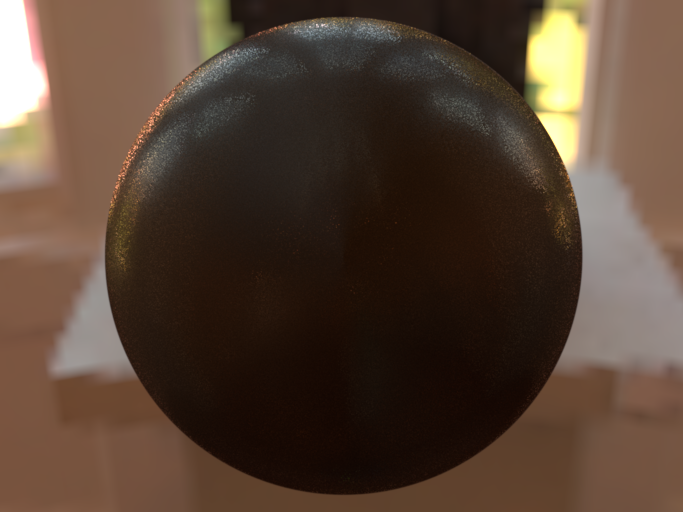}
    &\includegraphics[height=1.0in,width=0.40\linewidth]{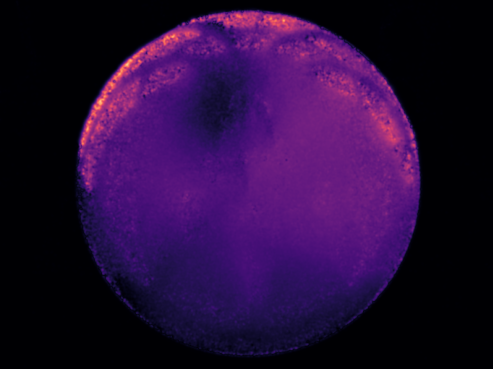} \\
    Novel Light with measurement&$\FLIP$ Error\\
    \end{tabularx}
\caption {Rendered images based on the brdf measurements from our method with the MERL Material:Alum-bronze. Note that the last image shows the $\FLIP$ error between the rendered image under novel lighting conditions using our measurements and the ground truth under the same novel lighting.}
\label{measure_merl}
\end{figure}

To evaluate the two components of our method—BRDF estimation (Sec.~\ref{net_brdf}) and adaptive measurements (Sec.~\ref{sampling})—we present the rendered images corresponding to each section using the Ward BRDF model and MERL material in Fig.~\ref{measure} and Fig.~\ref{measure_merl}.
\label{all_results}

In Fig.~\ref{measure},we show the rendered images under point lighting for visual comparison.The last image is rendered under an environment lighting to evaluate our measurements under novel lighting. We compare the rendered images obtained from ground truth BRDF values, Ward BRDF parameters estimated by the BRDF network, and adaptive sampling measurements—all of which appear visually almost identical. 

In Fig.~\ref{measure_merl},we use the Merl Material-Alum-bronze-to show the rendered images under point lighting. In the first row, we compare the images rendered using the ground truth measurements and our adaptive measurements, which appear almost visually identical. In the second row, we show the rendered image of our measurements under a novel environment light and the corresponding $\FLIP$ error image compared to the ground truth under the same environment lighting.

\subsection{Comparison}
Here,we compare our method with the state of the art method meta-learning brdf sampling method~\cite{liu_learning_2023-1}.Since Liu's method learns sample patterns for all materials, its performance does not improve with increased sample counts once highlights are captured. Additionally, it requires the implementation of a fixed sample count. In contrast, we derive adaptive sampling pattern for each specific input material, allowing the performance to progressively increase as more samples are added. We show the comparison of our method and Liu's method in Table~\ref{tab:results}, using the four overlapping test materials from each other's test datasets derived from the MERL dataset. And for quantitative comparison, we select results from Liu's method and our results with same sample numbers.

\begin{table*}[]%
\caption{Comparing Liu's method~\cite{liu_learning_2023-1},and our Adaptive sampler using all test Materials. Proposed Method in blue. Best values in bold.Note samples of Liu's method is in Rusinkiewicz parameterization($\phi_d,\theta_h,\theta_d$), while our sample's location is in spherical coordinates($\theta_{in},\phi_{in},\theta_{out},\phi_{out}$).}
\small
\label{tab:results}
\begin{center}
\begin{tabular*}{\textwidth}{@{\extracolsep{\fill}}l|l|ll|l|ll}
  \toprule
 \textbf {Test} &\textbf {Method}&Liu~\cite{liu_learning_2023-1}  &&\textbf {Method}& \textcolor{blue}{Image-based Adaptive}\\
  \cmidrule(lr){1-7}
  \textbf {Materials}&Samples Count&RMSE&PSNR&Samples Number & RMSE &PSNR
  \\
  \midrule\midrule[.1em]
Pink-fabric &32 &\textbf{0.0234}&\textbf{32.61}&$1\times8\times2\times2$&0.04&27.853\\ 
  &128&\textbf{0.0224}&\textbf{33.0}&$1\times8\times4\times4$&0.025&32.16 \\ 
   &256&\textbf{0.002}&\textbf{33.88}&$1\times8\times8\times4$&0.024&32.24\\
 &512 &\textbf{ 0.0211}&\textbf{33.53}&$1\times8\times8\times8$&0.0234&32.62\\
  
  \cmidrule(lr){1-7}
 Red-fabric2&32& 0.087&21.22&$1\times8\times2\times2$&\textbf{0.024}&\textbf{32.49}\\ 
  &128&0.08&21.6&$1\times8\times4\times4$&\textbf{0.01}&\textbf{39.72} \\ 
   &256&0.078&22.1&$1\times8\times8\times4$&\textbf{0.01}&\textbf{39.88}\\
 &512&0.08348&21.568&$1\times8\times8\times8$&\textbf{0.0098}&\textbf{40.17} \\
  
 \cmidrule(lr){1-7}
 Green-metallic-paint &32 & 0.1&20.16&$1\times8\times2\times2$&\textbf{0.056}&\textbf{25}\\ 
  &128 &0.0936&20.58&$1\times8\times4\times4$&\textbf{0.0246}&\textbf{32.17} \\ 
   &256&0.077&22.313&$1\times8\times8\times4$&\textbf{0.025}&\textbf{32.01}\\
  &512&0.081&21.82&$1\times8\times8\times8$&\textbf{0.02}&\textbf{33.94}\\
 
 \cmidrule(lr){1-7}
 White-diffuse-bball &32&0.0782&22.14&$1\times8\times2\times2$&\textbf{0.0221}&\textbf{30.36}\\ 
  &128&0.054&25.4&$1\times8\times4\times4$&\textbf{0.031}&\textbf{30.22} \\ 
   &256&0.032&30&$1\times8\times8\times4$&\textbf{0.031}&\textbf{30.054}\\
  &512&0.0324&29.79&$1\times8\times8\times8$&\textbf{0.029}&\textbf{30.65}\\
   
  \bottomrule
\end{tabular*}
\end{center}
\bigskip\centering
\vspace{-5pt}
\end{table*}%

The main results of the comparison is shown in Table~\ref{tab:results}.We set same sample count for all methods. Specifically, Liu’s method uses from 32 to 512 samples within the Rusinkiewicz parameterization, while our approach adopts a configuration of $1\times8$ incoming directions($\theta_{in},\phi_{in}$) and from $2\times2$ to $8\times8$ outgoing directions($\theta_{out},\phi_{out}$) in spherical coordinates. We observe that our method produces rendered images for all tests with high fidelity, whereas Liu’s method performs well only on specific materials. Except for pink fabric, our method outperforms Liu’s approach, as demonstrated by the results shown in Table~\ref{tab:results}.
\subsection{Evaluation on Different Materials}
\begin{figure*}[ht]
    \setlength{\tabcolsep}{0.01cm}
    \begin{tabularx}{1.00\linewidth}{ccccc}
      \textbf{Dark-red-paint}&\textbf{Color-changing-paint3}&\textbf{Alum-bronze}&\textbf{Dark-specular-fabric}&\textbf{Green-acrylic} \\
    \small\textbf{\rotatebox{90}{Single Point Light}}
        \includegraphics[height=0.9in,width=0.18\linewidth]{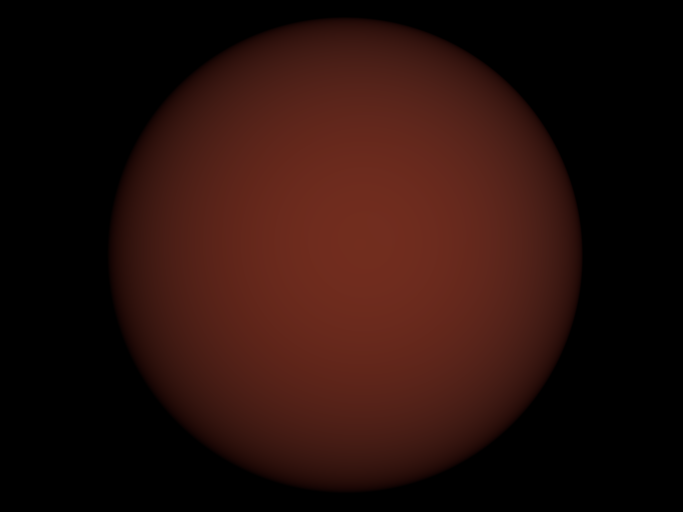}&\includegraphics[height=0.9in,width=0.18\linewidth]{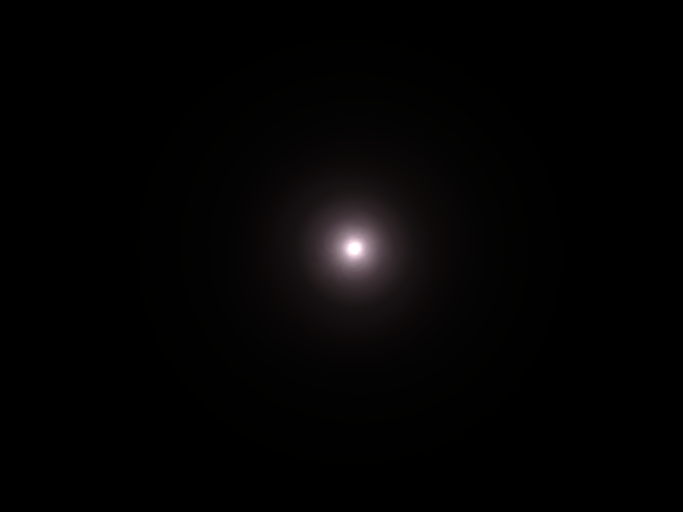}&\includegraphics[height=0.9in,width=0.18\linewidth]{Figures/merl_measurements/alum-bronze/adaptive/4_16_32_32_adapative.png}&\includegraphics[height=0.9in,width=0.18\linewidth]{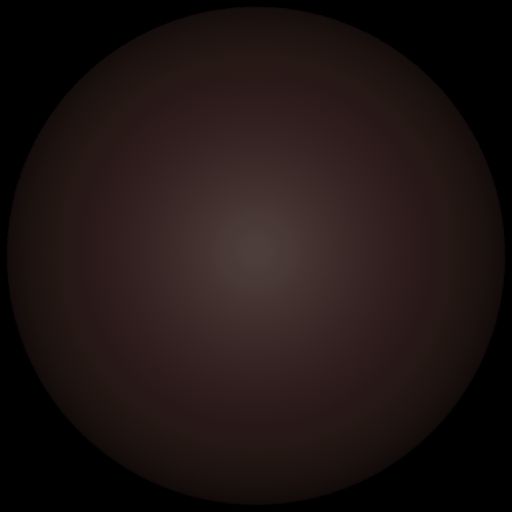}&\includegraphics[height=0.9in,width=0.18\linewidth]{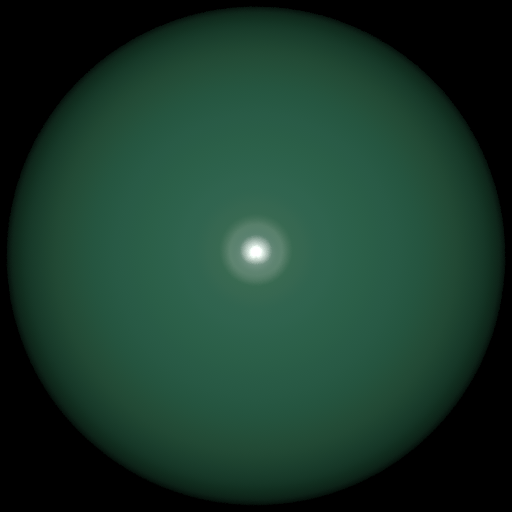}
        
        \\
       \small\textbf{\rotatebox{90}{Environment Light}}
        \includegraphics[height=0.9in,width=0.18\linewidth]{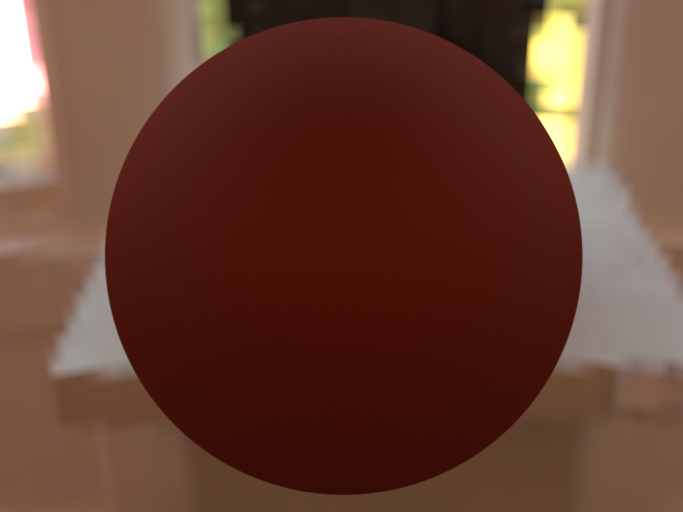}&\includegraphics[height=0.9in,width=0.18\linewidth]{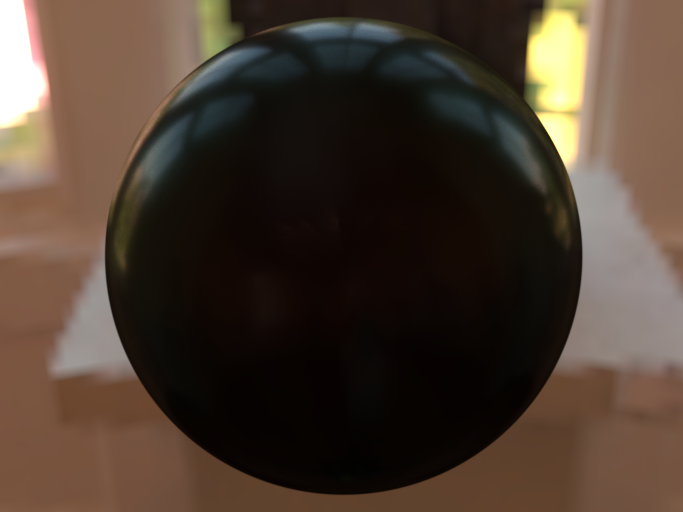}&\includegraphics[height=0.9in,width=0.18\linewidth]{Figures/merl_measurements/alum-bronze/adaptive/4_16_32_32_adapative_env.png}&\includegraphics[height=0.9in,width=0.18\linewidth]{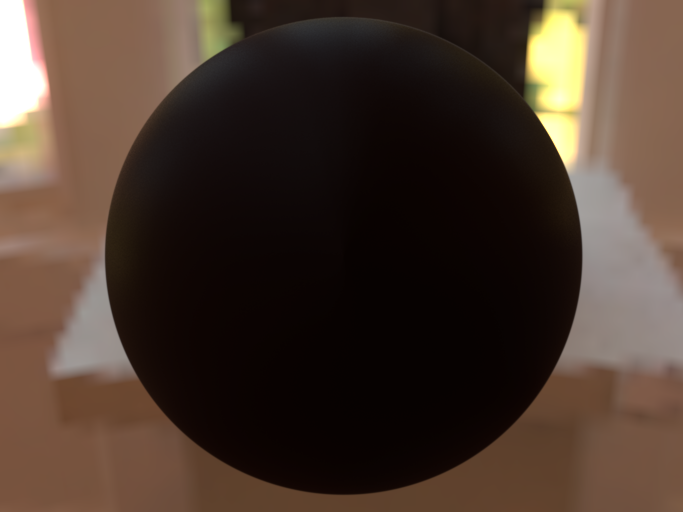}&\includegraphics[height=0.9in,width=0.18\linewidth]{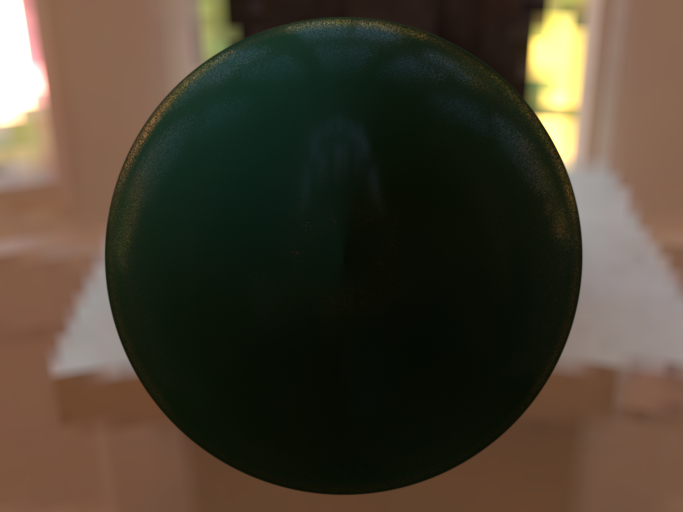} \\   
       \small\textbf{\rotatebox{90}{Ground Truth}}
        \includegraphics[height=0.9in,width=0.18\linewidth]{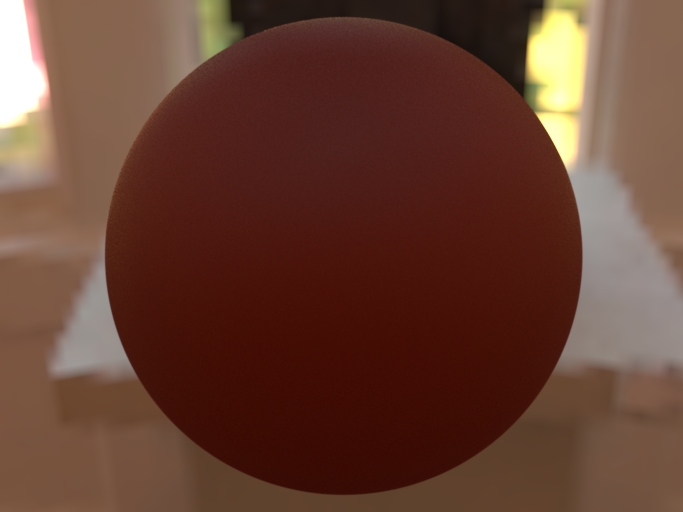}&\includegraphics[height=0.9in,width=0.18\linewidth]{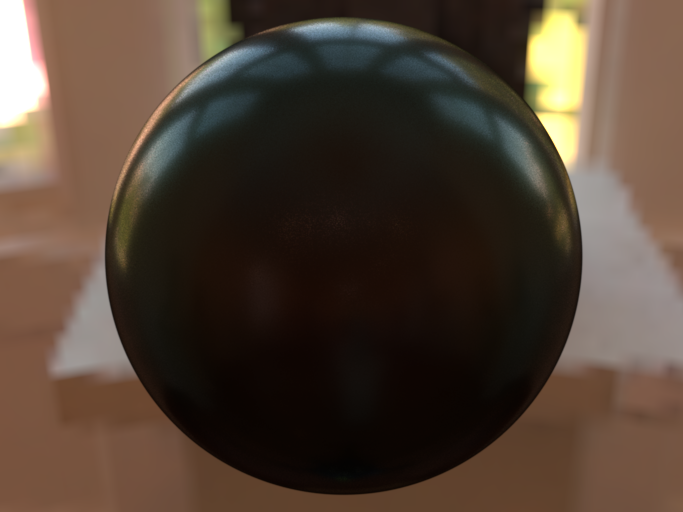}&\includegraphics[height=0.9in,width=0.18\linewidth]{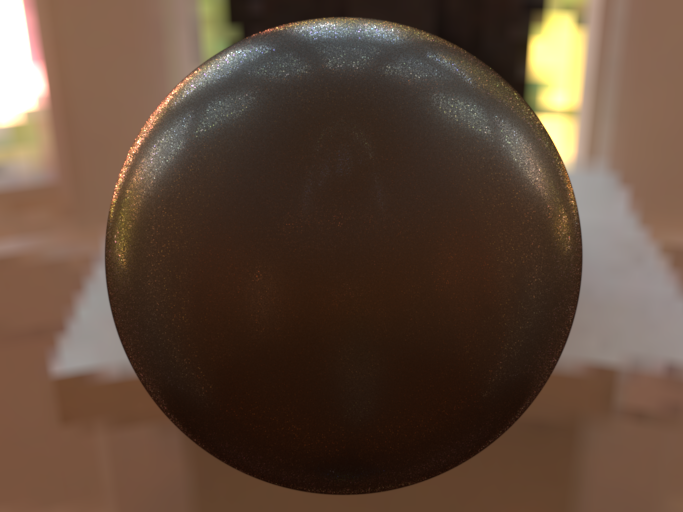}&\includegraphics[height=0.9in,width=0.18\linewidth]{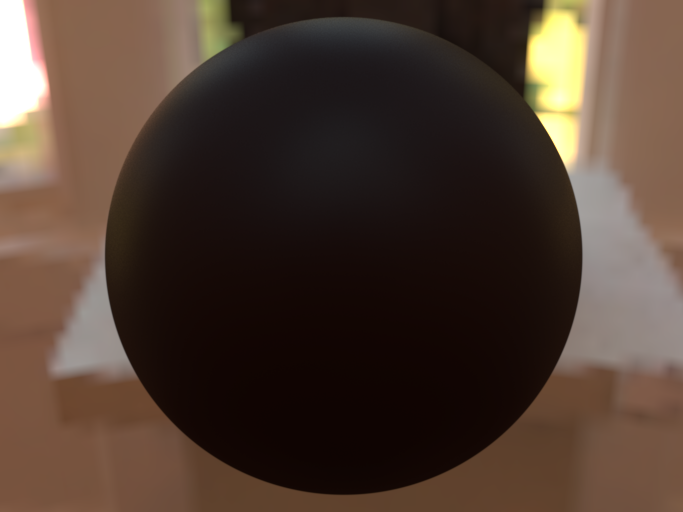} &\includegraphics[height=0.9in,width=0.18\linewidth]{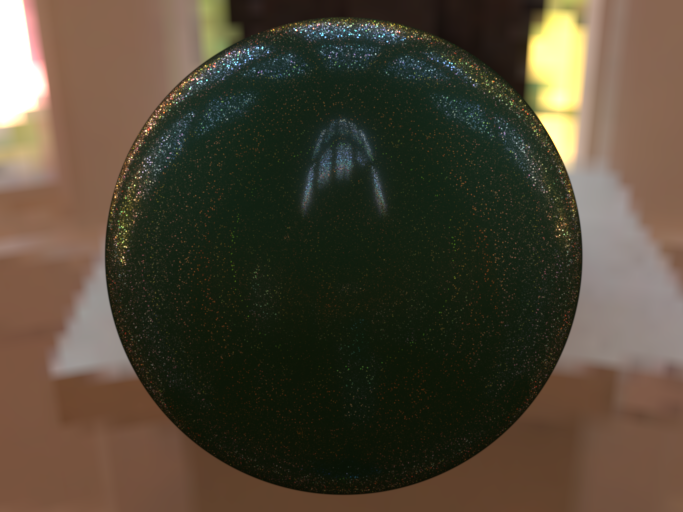}\\
       
        \small\textbf{\rotatebox{90}{ \FLIP Error}}\includegraphics[height=0.9in,width=0.18\linewidth]{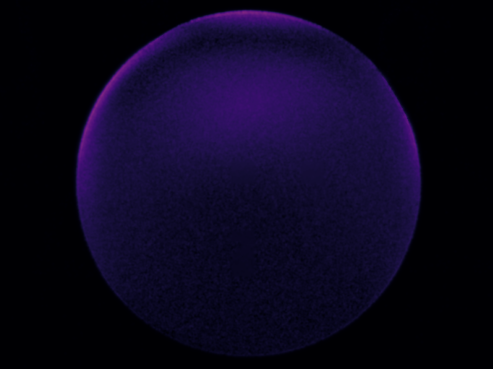}&\includegraphics[height=0.9in,width=0.18\linewidth]{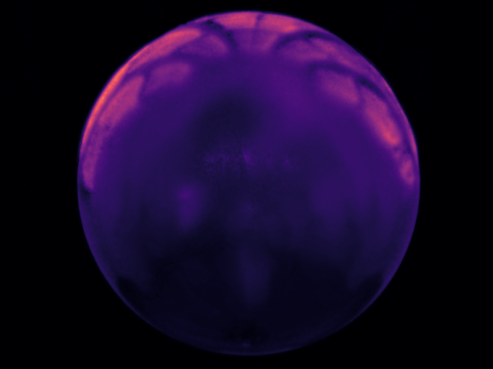}&\includegraphics[height=0.9in,width=0.18\linewidth]{Figures/merl_measurements/alum-bronze/adaptive/flip_error_adap.png}&\includegraphics[height=0.9in,width=0.18\linewidth]{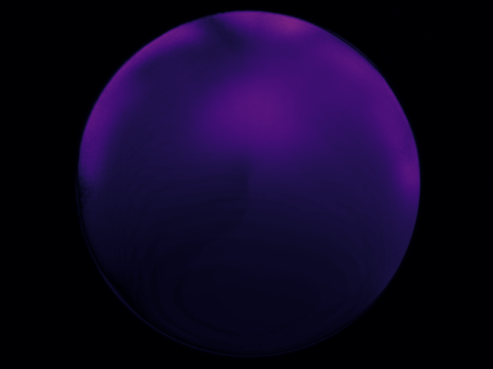}&\includegraphics[height=0.9in,width=0.18\linewidth]{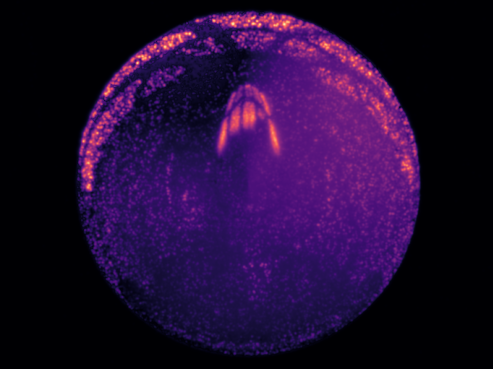}\\ 
         $8\times8$&$16\times16$ &$16\times16$&$14\times14$&$32\times32$\\
        \small\textbf{\rotatebox{90}{ Performance Plot}}\includegraphics[height=0.9in,width=0.18\linewidth]{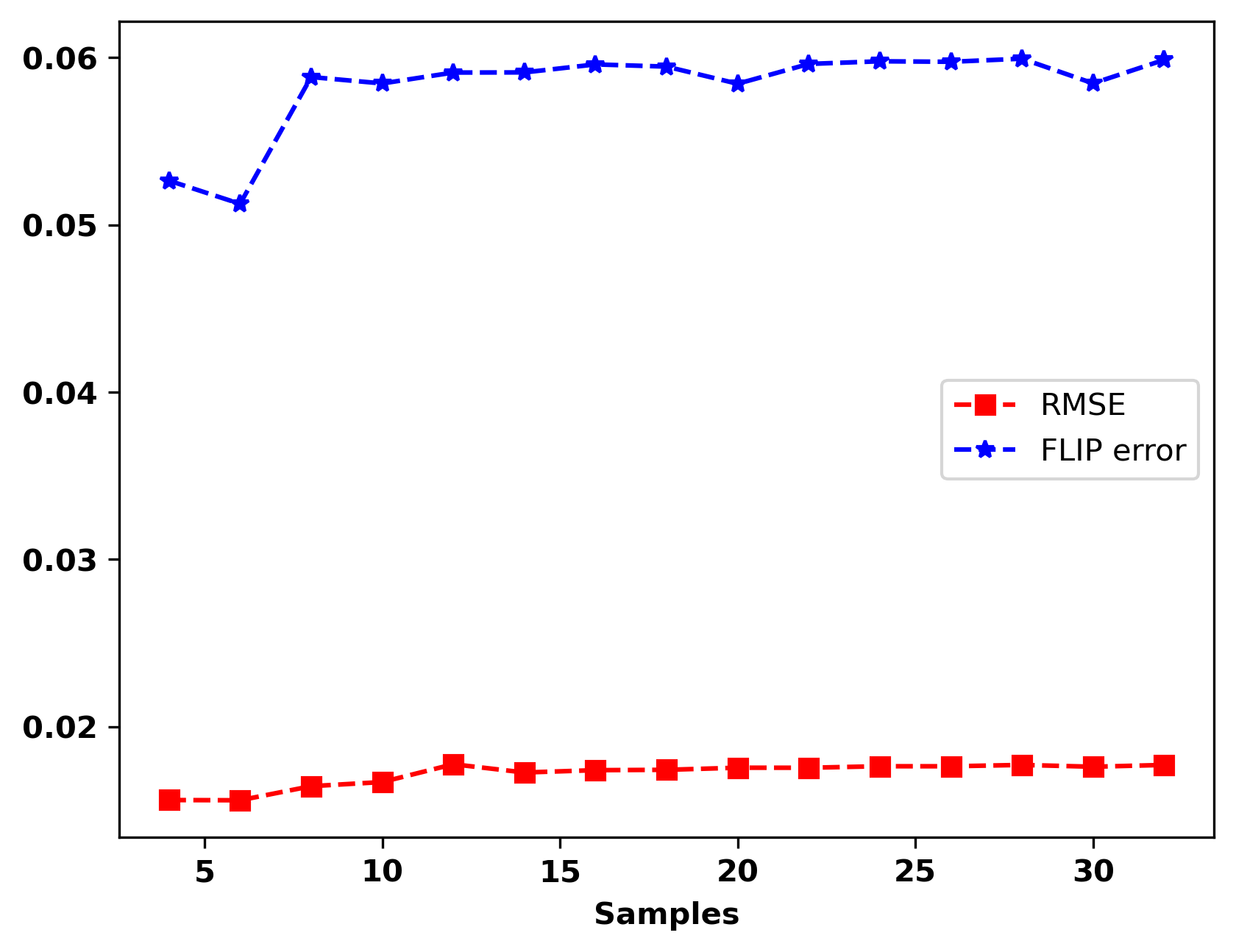}&\includegraphics[height=0.9in,width=0.18\linewidth]{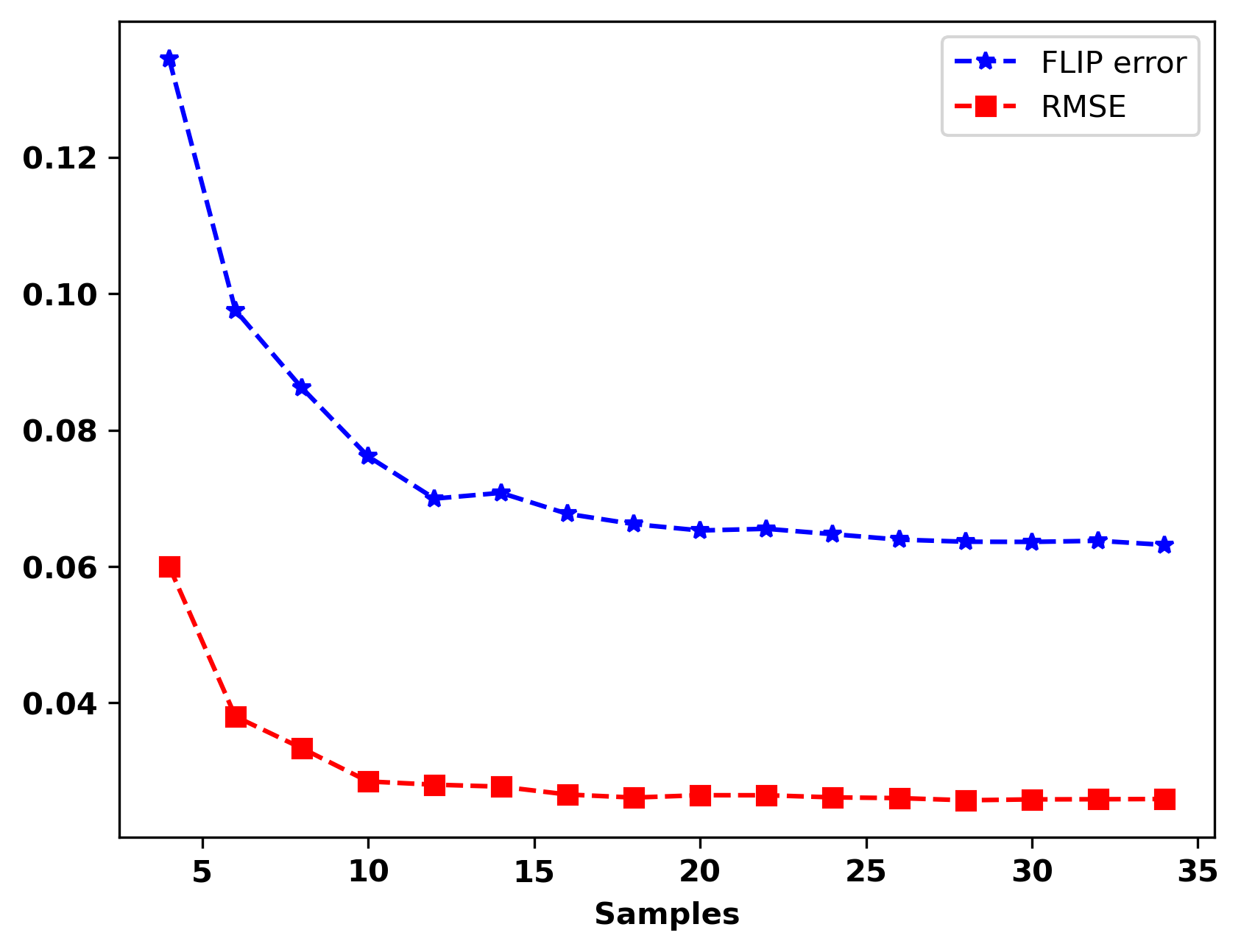}&\includegraphics[height=0.9in,width=0.18\linewidth]{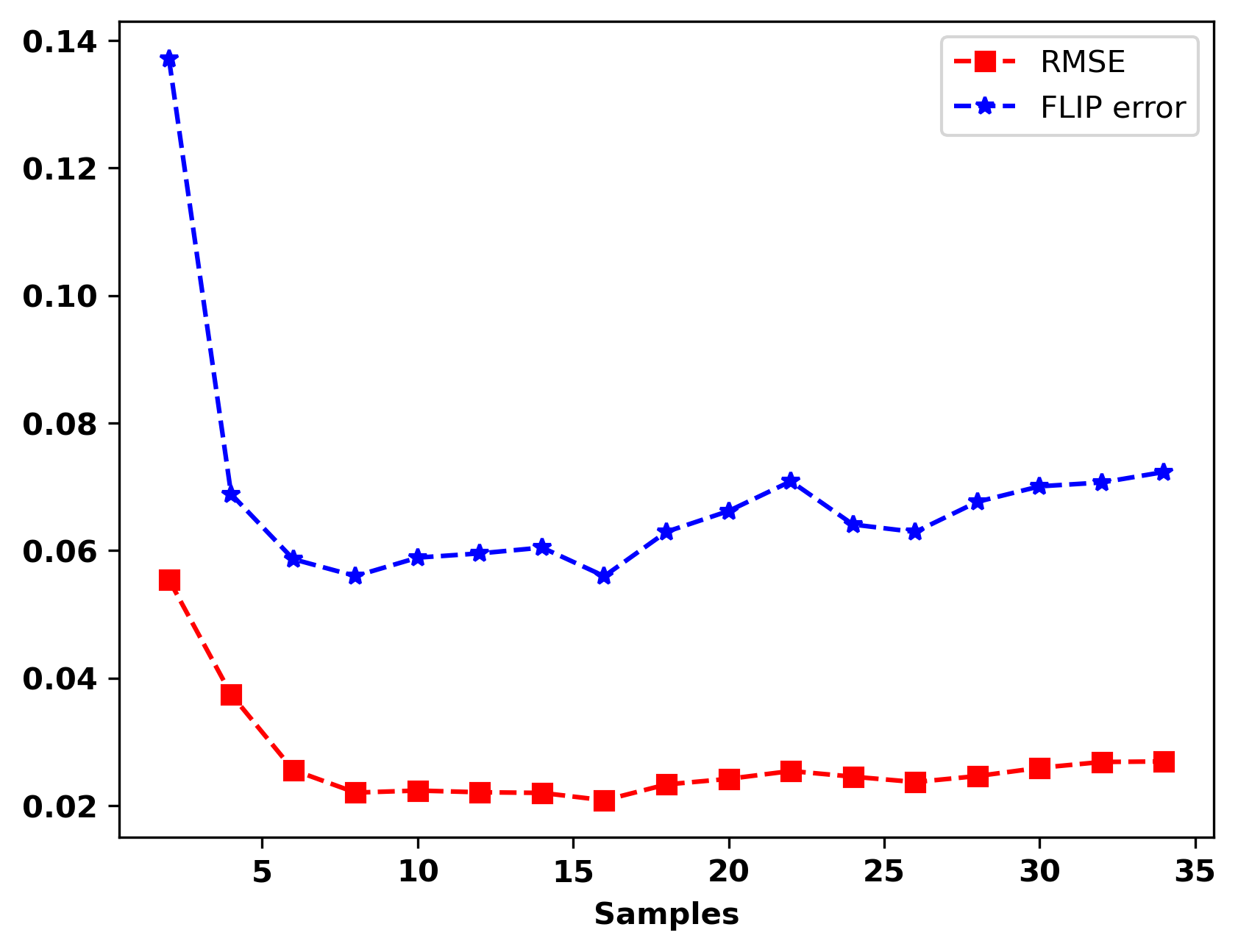}&\includegraphics[height=0.9in,width=0.18\linewidth]{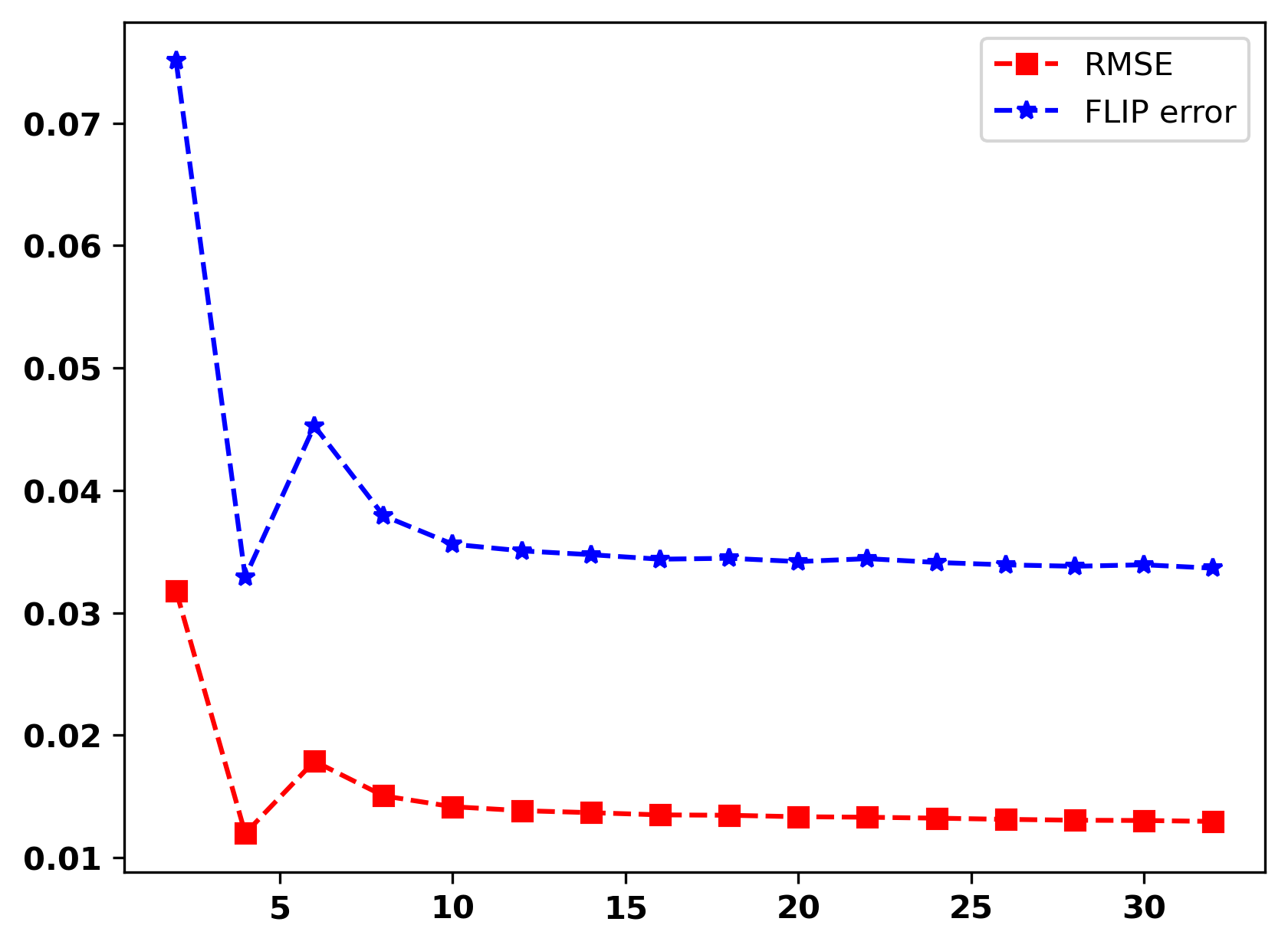}&\includegraphics[height=0.9in,width=0.18\linewidth]{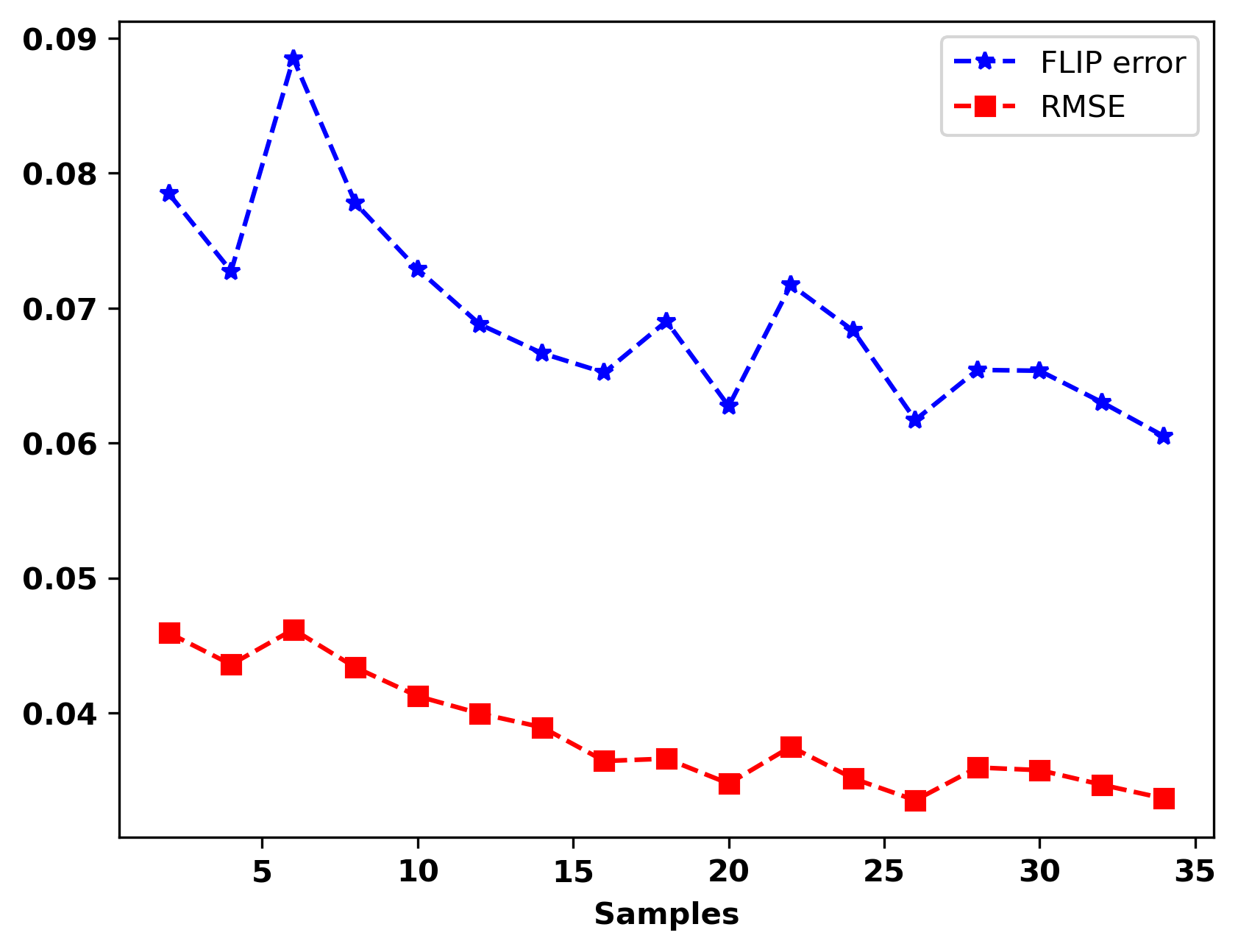}\\
       
        \end{tabularx}
    \caption{ Rendered Images of Five Different Materials from the MERL Dataset.The first row shows our measurements rendered under single-point lighting. The second row shows our measurements rendered under environmental lighting. The third row shows the ground truth rendered under the same environmental lighting conditions. The fourth row shows the $\FLIP$ error images between the second and third rows. The final row depicts the plot of metric values  with the sample numbers for each material. }
    \label{merl_adaptive}
\end{figure*}

We show more visual results in Fig.~\ref{merl_adaptive} with five different materials.We observe that diffuse materials achieve high-quality results with fewer samples, whereas highly specular materials require a larger number of samples to produce good outcomes.The first row and second row are the rendered images using our adaptive measurement method under point lighting and environment lighting, respectively. The fourth rows of Fig.~\ref{merl_adaptive} shows the $\FLIP$ Error image comparing the second row (rendered images from our method) with the third row (ground truth), where specular materials generally exhibit higher errors.The last row presents the plot of the performance metrics based on sample numbers, illustrating how the sample count is selected for each material as described in section~\ref{sample_num}. Additional more rendered images of test materials are available in the supplementary materials.

\section{Conclusion}
We propose an image-based adaptive BRDF sampling method that significantly accelerates BRDF measurements while maintaining high accuracy and fidelity through a lightweight neural network. We validate our approach using both the MERL dataset and the Ward BRDF model. Additionally, we compare our method against the state-of-the-art method by Liu et al~\cite{liu_learning_2023}. Our method outperforms in comparison.

\section{Disscusion and Limitation}
Moreover, additional variants can be explored in this approach.

\textbf{Sample method} Normalizing Flows~\cite{muller_neural_2019} is a possible alternative for our adaptive sampler as it can generate samples in inverse and forward way.Exploring the integration of Normalizing Flows into our sampling strategy could be a valuable direction for future research.

\textbf{BRDF analytic model} We evaluated the Ward BRDF model and the Microfacet model. We believe that other BRDF models can also be effectively incorporated into our approach.

\textbf{Training and Test Dataset} For our experiments, we use spherical geometry to produce rendered images for both the training and test datasets. We anticipate that replacing the sphere with a planar geometry could be feasible within our method.

\textbf{Limitation} We see that the BRDF estimation network is sensitive to variations in light intensity. Addressing this sensitivity to enhance the network’s robustness against different lighting conditions represents an important area for future work.


\bibliographystyle{apalike}
{\small
\bibliography{main}}

\end{document}